\documentclass[twocolumn]{29onr_art}
\usepackage{natbib}
\usepackage{ulem}
\usepackage{graphicx}
\usepackage{fixltx2e}
\usepackage{amsmath,float}
\usepackage{graphicx}
\usepackage[english]{babel}
\usepackage{epsfig}
\usepackage{color}
\usepackage[colorlinks=true,
            urlcolor=blue,
            linkcolor=docgreen,
            citecolor=docgreen,
            bookmarks=true,
            pdftitle={The Impact of a Deep-Water Plunging Breaker},
            pdfauthor={Christine Ikeda},
            pdfsubject={Abstract Submitted for the 29th Naval Hydrodynamics Symposium},
            pdfkeywords={Breaking Waves, Wave Impact, Air Entrainment, Numerical Simulations},
	    bookmarksopen=false,
            pdfpagemode=None]{hyperref}

\definecolor{docgreen}{rgb}{0,.5,0}

\pdfoutput = 1

\begin{document}

\title{The Impact of a Plunging Breaker on a Partially Submerged Cube}

\author{
Christine Ikeda$^1$, Thomas T. O'Shea$^2$, Kyle A. Brucker$^2$, \\
David A. Drazen$^3$, Douglas G. Dommermuth$^2$, Thomas Fu$^3$, \\
Anne M. Fullerton$^3$, and James H. Duncan$^1$}

\affiliation{\small $^1$University of Maryland, College Park, Maryland, U. S. A. \\
$^2$Science Applications International Corporation, San Diego, California, U.S.A. \\
$^3$Naval Surface Warfare Center, Carderock, Maryland, U.S.A.}

\maketitle

\begin{abstract}
The impact of a plunging breaking wave (wavelength $\approx
1.3$~m) on a rigidly mounted rigid cube structure (dimension 0.31~m) that
is partially submerged is explored through experiments and numerical
calculations.  The experiments are carried out in a wave tank and the
breaker is generated with a mechanical wave maker using a dispersive
focusing technique.  The water-surface profile upstream of the front
face of the cube and in its vertical centerplane is measured using a
cinematic laser-induced fluorescence technique.  The three-dimensional
flow in the wave tank is simulated directly using the Numerical Flow
Analysis (NFA) code.  The experiments and the calculations are used to
explore the details of the wave-impact process and, in particular, the
formation of the high-speed vertical jet that is found on the front
face of the cube under some impact conditions.

\end{abstract}

\section{Introduction}

The impact of large-amplitude waves on structures has been investigated by a number of researchers.  The literature on wave impact on vertical walls, typically mounted to the bottom boundary in shallow water, has been reviewed by \cite{Peregrine:2003}.  Wave breaking is a strongly nonlinear transient two--phase process that is not easily modeled with theoretical or numerical treatments and makes well-controlled repeatable experiments very difficult. In fact, it was not until the 1970s that accurate numerical modeling of the overturning of a water wave has been achieved by researchers using potential flow, boundary-element calculations, see for example \cite{long1976}, \cite{cooker1990computations} and  \cite{dold1986efficient}.

\begin{figure*}
\renewcommand{\baselinestretch}{1}
\begin{center}
\includegraphics[trim=0 0.0in 0 0.00in,clip=true,scale=0.5]{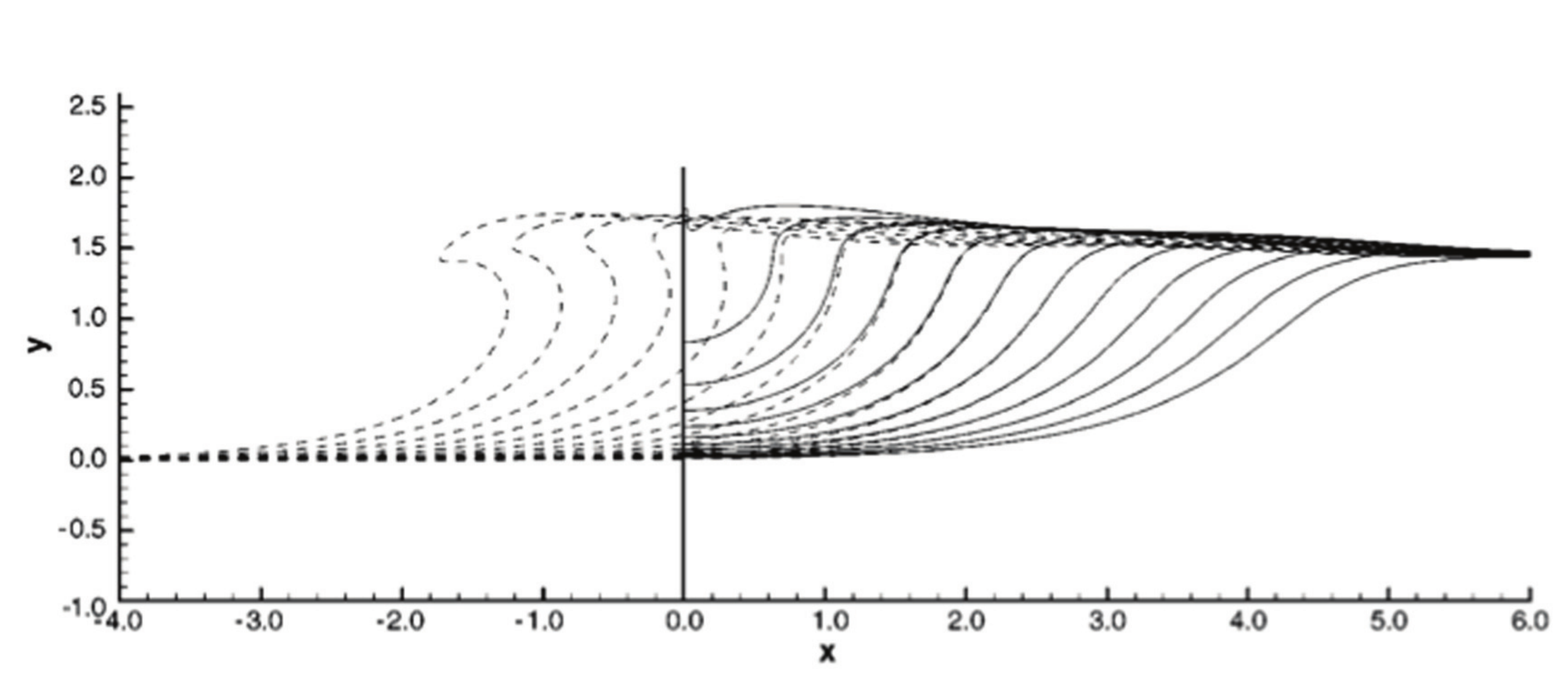}
\end{center}
\vspace*{-0.3in} \caption{The dashed lines show the profiles of a plunging breaking wave without the influence of a wall. The solid lines show the same plunging breaking wave with a wall located where the dashed line profiles appear to be most vertical. Due to the presence of the wall, the water below the crest is not allowed to move forward in the streamwise direction, and moves upward along the wall. The cavity appears to close as it approaches the wall. \cite{Peregrine:2003} calls this phenomenon ``flip-through.''} \label{fig:PereWaveApproach}
\renewcommand{\baselinestretch}{2}
\end{figure*}

Without the presence of a wall, a plunging breaking wave can occur when, due to various nonlinear effects, the wave steepens rapidly forming a jet that curls over and plunges back into the front face of the wave. The dashed lines in Fig.~\ref{fig:PereWaveApproach} (taken from Peregrine (2003)) show profiles from numerical calculations of a plunging wave that breaks in open water due to nonlinear effects caused by propagation in shallow water of constant depth.  If a bottom-mounted vertical wall is placed in the computational domain, the breaking process can change dramatically depending on the location of the wall.  The water surface profiles shown as solid lines in Fig.~\ref{fig:PereWaveApproach} were obtain with the wall placed at the streamwise location where the wave front without the presence of the wall is the most vertical.  As can be seen in the figure, the water is pushed upward by the wall and forms a vertical jet (not shown in the figure).  The wave impact pressure is strongly affected by the water depth, the submergence of the wall and the shape evolution of the wave as it approaches the wall.  \cite{CHAN:1988}, who conducted an extensive experimental study considering the impact of deepwater waves on a vertical wall that spanned the width of the tank and extended to the tank bottom, describe three different regions of wave impact: 1) a region of wall positions where a strong vertical jet is formed during impact  and high impact pressures are observed (called ``flip-through'' by \cite{Peregrine:2003}, the beginning phases of which are shown in Fig.~\ref{fig:PereWaveApproach}); 2) a transition region where no significant impact pressures are observed; and 3) a region where the crest has already plunged into the fluid before impacting the wall and a double-peak impact pressure is observed due to a secondary jet impacting the wall ahead of the crest.  \cite{CHAN:1988} found large variability in the observed pressure record in repeated runs with flip-through behavior indicating that the details of this phenomenon might be sensitive to residual water motions from previous experimental runs.  However, both \cite{Peregrine:2003} and \cite{CHAN:1988} observed ``flip-through'' behavior even though the incident wave generation was very different.

% The flip-through phenomena seems to be sensitive to 
% High pressures and violent upward jets can also be caused by ``flip-through,'' which is the phenomenon of the smooth water-surface focusing to a point. ``Flip-through'' is caused by the presence of the wall not allowing the water to move in the streamwise direction pushing the water upward in the vertical direction [\cite{zeff2000singularity, Peregrine:2003}].  

% The ``Flip-through'' can be thought of as an upper-bound for impact pressures, because in reality, the free-surface or wall is not likely to be smooth enough to cause the focusing to occur outside of a laboratory environment. The high localized pressures can be minimized by small disturbances such as surface roughness on the wall or small surface waves. Only the first wave in an experimental facility will be smooth, and therefore have the largest pressures. Even though the surface is smooth, researchers such as \cite{CHAN:1988} found large variability in the observed pressure record. 

The pressure on a wall during wave impact can be attributed to both the hydrodynamics of the wave impact that causes high impulsive pressures and also the dynamics of entrained air that can cause higher amplitude pressures and pressure oscillations after the impact event. During an impact event, air can be entrained when the wave breaks before impacting the wall or when the jet impacts the wall before breaking thus entrapping a pocket of air.  The compressibility of the air-water mixture due to air entrainment causes problems in comparing laboratory experiments to full scale.  \cite{lamarre1992instrumentation} and \cite{ bullock1999characteristics} measured a wide range of air-volume fractions in breakers generated in the laboratory and those found in the field.  In the experiments of \cite{CHAN:1988}, variabilities of the impact pressure peak from run to run (for identical initial conditions) are attributed to the randomness in the bubble oscillations, also making simple scaling unjustifiable. \cite{bagnold1939interim}, \cite{chan1994mechanics}, \cite{CHAN:1988}, \cite{Peregrine:2003} argue that the pressure impulse of the wave impact is a more consistent metric compared with the peak pressure. The pressure impulse is simply the integral of the pressure over time for the duration of the impact.

\cite{Peregrine1996effect} used a simple equation of state for water and air mixtures, and found that the peak impact pressure on the wall is  reduced by increasing the air fraction. \cite{Peregrine:2003} explains that the air pocket entrapped on the wall takes time to be compressed and then re-expand, which causes an increase in the time duration of the pressure peak. The interaction with the air pocket acts as a spring causing the water to rebound or ``bounce-back'' similar to an elastic collision. Peregrine's analysis  led to a increased pressure impulse on  the wall, which was shown experimentally by \cite{wood2000air}. When there is a ``flip-through'' wave impact , which by definition forms no air pocket, some of the energy from the incoming wave is converted into the upward jet. In cases where an air pocket is entrapped by the incoming wave, energy that would have gone into the upward jet is instead used to break up the air pocket. This would then lead to a reduction in the jet height when air pockets are entrapped.

A number of wave impact studies were directed toward assessing the effectiveness of breakwaters for protection against damage due to tsunamis.  \cite{Oshnack09} conducted a large-scale experimental study of the effectiveness of small seawalls. \cite{lukkunaprasit2008building} found that low retaining walls (1~m high) were effective at dissipating tsunami energy. \cite{dalrymple2005lessons} discussed the idea of the upward splash of the wave impact deflecting the wave momentum upward and thus reducing damage to structures on the other side of the seawall. They observed that there was increased damage after a tsunami at positions directly behind the pedestrian openings in the seawall, suggesting that the seawalls do in fact deflect the wave energy and reduce damage.

% This is the work compared with Oregon State University
% Two-dimensional plunging breakers were produced by a dispersive focusing method (\cite{Rapp:1990}) and the position of the wall relative to the breaker location without the wall was varied.  The wave impact and peak pressures on the wall were found to vary with the wall position and peak pressures ranging from $3\rho C^2$ to $10\rho C^2$ were found, where $\rho$ is the density of water and $C$ is the wave phase speed. 

\cite{fu08} and \cite{nfa2} investigated wave impact loading on the front face of a box with dimensions of 0.305~m by 0.305~m by 1.52~m.  The box was mounted in the water-surface in the model basin at the Naval Surface Warfare Center in Carderock, Maryland (dimensions 6.4~m wide, 3~m deep and 514~m long) with the long axis of the box horizontal and parallel to the wave crests and the front face of the box at angles of $0^\circ$, $45^\circ$ and $-45^\circ$ relative to horizontal.  The center section of the structure was a water-tight cube and its front face was instrumented with 11 pressure sensors and 9 flexible panels that were in turn instrumented with strain gauges.   Two-dimensional breakers with wavelengths of 6.1~m and 9.1~m were produced by a dispersive focusing method (\cite{Rapp:1990}).  Due to the nature of the pneumatic wave maker in the model basin, it was not possible to produce highly repeatable waves. The Numerical Flow Analysis (NFA) code was used to make comparisons to the laboratory measurements. The predicted range of peak loads were in agreement with the laboratory measurements.

In the present work, the center cube section from the model used in the above-described experiments was placed in the wave tank at the University of Maryland and a series of experiments with simultaneous wave crest shape, impact-pressure and impact-force measurements is underway.  Two-dimensional plunging breakers are  produced by a dispersive focusing method similar to \cite{Rapp:1990}, and the position of the cube relative to the breaker location without the cube is varied. The wave profiles are measured using a cinematic Laser-Induced Fluorescence (LIF) technique. It is important to note that the cube does not span the entire width of the wave tank thus creating a three-dimensional wave impact; therefore, the physics of this experiment are different from the works cited previously where a two-dimensional problem is discussed and a wall was mounted on the bottom of the tank. It is expected from \cite{wood1998two} and \cite{Peregrine:2003}, that the pressures and forces will be reduced compared to the two-dimensional studies.  These experiments are being simulated with the CFD code NFA.  The combined experiments and numerical analysis are being used to explore the physics of the wave-structure interaction problem. The results presented herein consist of a comparison of the wave profiles between the experiments and the NFA simulations for one cube position along with experimental measurements of the position and velocity of the contact-point between the free surface and the vertical center line of the cube.  In addition to this work, a set of calculations describing the water surface profiles and impact pressures due to a shoaling wave impinging on a vertical wall are presented and compared to published experimental results.

\section{Experimental Details}

The experiments are performed in a wave tank that is 14.8~m long,
1.5~m wide and 2.2~m deep with a water depth of 0.91~m.  The waves are
generated by a programmable wave maker consisting of a vertically
oscillating wedge that is located at one end of the tank.   The front
face of the wedge is tilted forward by 30$^\circ$ and the back face of
the wedge is vertical and very close to the end wall of the tank.  
The wave-maker is controlled by a computer-based
feedback system and its motions are repeatable to $\pm$0.1\% in
amplitude from run to run.  These highly repeatable wave maker motions
result in highly repeatable breaking waves.

\begin{figure*}
\begin{center}
\includegraphics[trim=0 0.0in 0 0.00in,clip=true,scale=0.75]{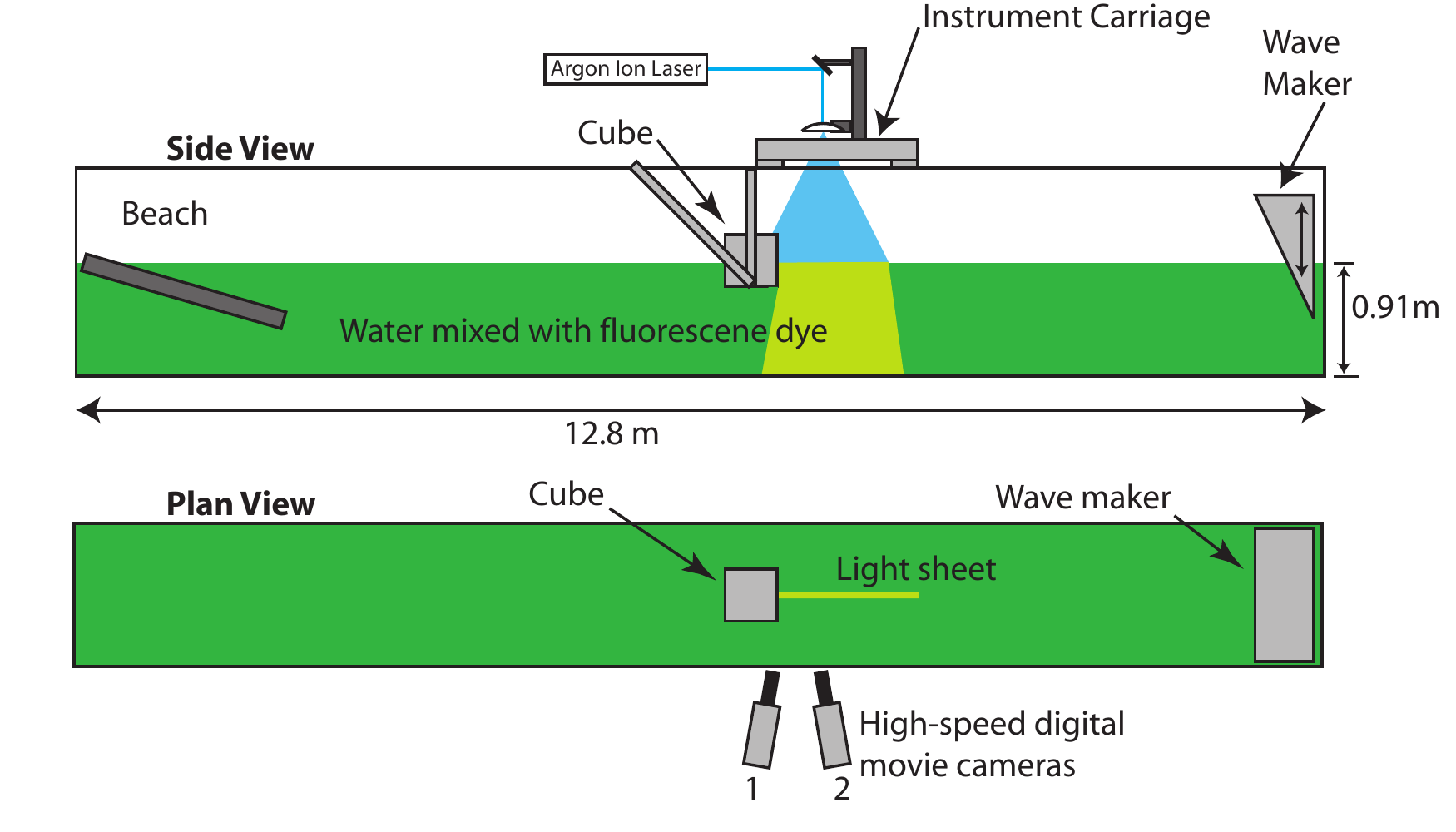}
\end{center}
\vspace*{-0.3in} \caption{A schematic showing the tank, the
wave maker, the cube and the lighting setup from the cinematic LIF
measurements of the water-surface profiles.} \label{fig:tank_pic}
\end{figure*}

The breaking waves were generated by a dispersive focusing method similar to that described
in  \cite{Rapp:1990}.    In this method, a packet of waves
with varying frequency is generated in a manner such that the packet
converges as it travels along the tank.   This convergence causes the 
amplitudes of the remaining waves in the packet  to increase. Eventually
a breaking wave is formed  if the initial wave amplitudes are large
enough.   Linear deep-water wave theory is used to compute a suitable
motion for the wavemaker though the resulting wave motion is highly
nonlinear when the packet converges. The wave packet consists of the sum
of $N$ sinusoidal components and the wavemaker motion to produce
these waves is given by
\begin{eqnarray}
z_w & = & w(t) \frac{2\pi}{N}A  \sum_{i=1}^{N}  \frac{1}{k_i} \times \nonumber \\
& = & 
\cos(x_b (\frac{\omega_i}{\bar{c}} -
k_i)-\omega_i t + \frac{\pi}{2}), 
\end{eqnarray}
where $w$ is a window function which is described below, 
$A$ is an adjustable constant called the wavemaker amplitude, 
$x_b$ is the horizontal position of the breaking event (by linear
theory) measured from
the back of the wedge (for purposes of the experiment), $t$ is time, 
$k_i$ and $\omega_i$ are, respectively, the wavenumber 
and frequency of each of the $i=1$ to $N$ wave components 
($\omega = 2 \pi f$
where $f$ is the frequency in cycles per second), and 
$\bar{c}$ is the average of the
group velocities ($c_i=0.5\omega_i/k_i$) of the $N$ components.
The frequencies are equally spaced, $\omega_{i+1}=\omega_i+\Delta
\omega$, where $\Delta \omega$ is a constant.
The window function was chosen to give the wedge zero motion
at times when the summation of components resulted in  only a very small
motion:
\begin{eqnarray}
w(t) & = &  0.25(\tanh(\beta\bar{\omega}(t-t_1))+1) \times \nonumber \\ 
& & (1-\tanh(\beta\bar{\omega}(t-t_2))),
\end{eqnarray}
where $\beta$ is a constant that determines the rise rate of the 
window function, chosen as $5.0$, and
$\bar{\omega}$ is the average of the $N$ frequencies $\omega_i$.  The
window function is nearly equal to $1.0$ for most of the time between
$t= t_1$ and $t=t_2$ and is zero at other times.  The times $t_1$ and
$t_2$ were chosen to allow the lowest and highest frequency components
($i=1$ and $i=N$, respectively) to be generated and to 
travel to position $x_b$:
\begin{eqnarray}
t_1 & = & x_b\left(1/\bar{c} - 1/c_N\right) \\
t_2 & = & x_b\left(1/\bar{c} - 1/c_1\right) 
\end{eqnarray}

The parameters controlling the wave maker motion were adjusted by
trial and error to produce a plunging breaker from a wave packet with
an average frequency, $f_0=1.15$~Hz.  For this breaker, $N=32$, $h/\lambda_0 =
0.35792$ (where $h$ is the vertical distance between the mean water
level and the vertex of the wedge and $\lambda_0 = 2\pi
g/\bar{\omega}^2$, where $g$ is gravitational acceleration),
$H/\lambda_0=0.486$ ($H$ is the mean water depth in the tank),
$x_b/\lambda_0 = 10.0$, $N\Delta \omega/\bar{\omega} = 0.77$, and
$A/\lambda_0 = 0.074$.

The cube is made from 12.7-mm-thick aluminum plates and the final
structure is 30.48~cm on each side.  The cube is attached to an
aluminum frame which is in turn attached the the tops of the tank walls
as shown schematically in Figure~\ref{fig:tank_pic}.  The cube is
located at various horizontal positions in the tank relative to the
location of the wave breaking location (as determined when the cube is removed from
the tank). In the vertical direction, the cube is positioned with its
horizontal middle plane in the plane of the undisturbed water-surface.

Water-surface profile observations were taken with two high-speed digital
movie cameras (Phantom model V640, Vision Research, Inc.) using diffuse
white-light illumination to determine the qualitative behavior of the
impact and
using Laser-Induced Fluorescence (LIF) for quantitative measurements.  The
cameras are located outside the clear plastic tank walls and view the
wave impact from the side, just in front of the cube.  One camera
looks slightly upstream and the other slightly downstream, see
Figure~\ref{fig:tank_pic}. The
illumination for the LIF measurements is a vertically oriented light
sheet from a six-watt argon-ion laser as shown in
Figure~\ref{fig:tank_pic}.  It is estimated that the surface height at any location and any instant in time is measured to an accuracy of 
$\pm 0.75$~mm.

\section{Numerical Setup}

A Volume of Fluid (VOF) and  a Boundary Integral Equation Method (BIEM) are used to model the wave tanks.    The VOF formulation is based on the Numerical Flow Analysis (NFA) method.   The BIEM formulation is based on a formulation that had been compared to experimental measurements of plunging breaking waves \citep{dommermuth88}.  The NFA and BIEM methods are used to simulate waves generated by a wedge wavemaker impacting on a cube.  In the case of BIEM, the cube is not modeled.  The NFA method is also used to simulate a soliton generated by a piston wavemaker breaking on a beach and impacting a breakwater.   Highlights of the numerical methods are provided in the next two sections.

\subsection{NFA Formulation}

\begin{figure*}
\begin{center}
\begin{tabular}{c}
\includegraphics[width=0.45\linewidth]{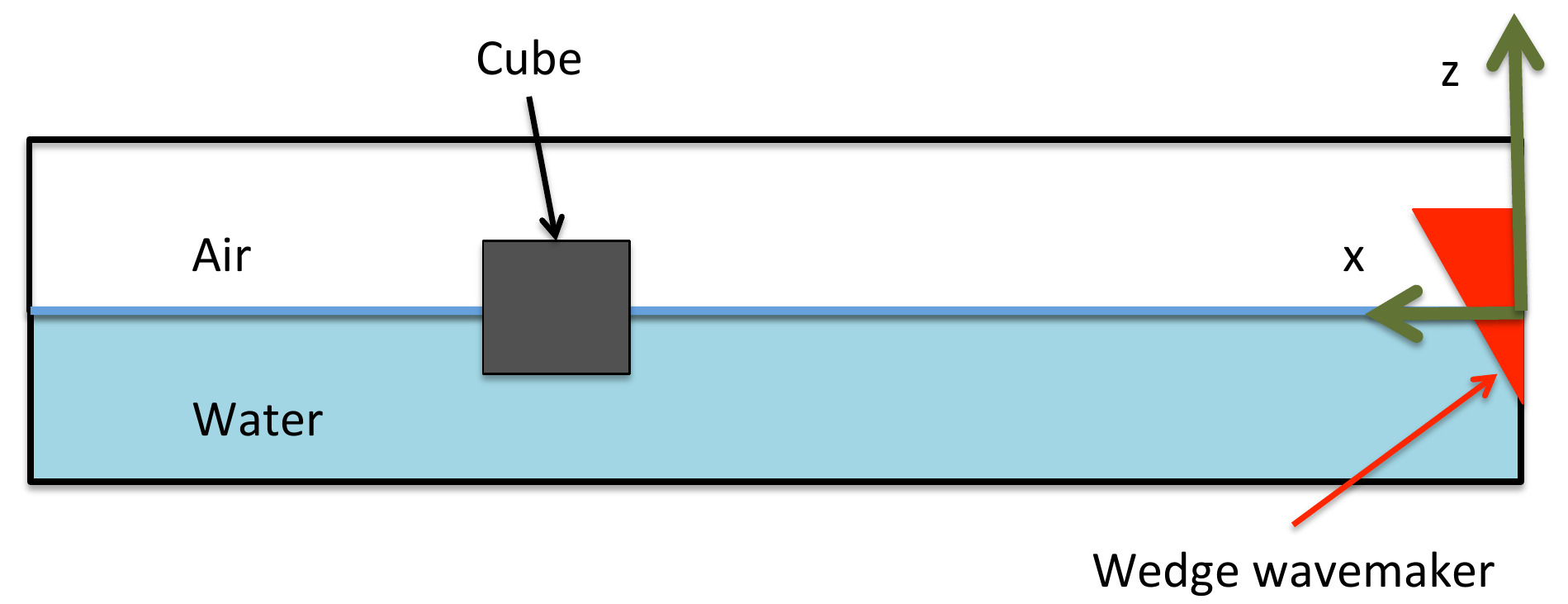}   \\
(a) \\
\includegraphics[width=0.5\linewidth]{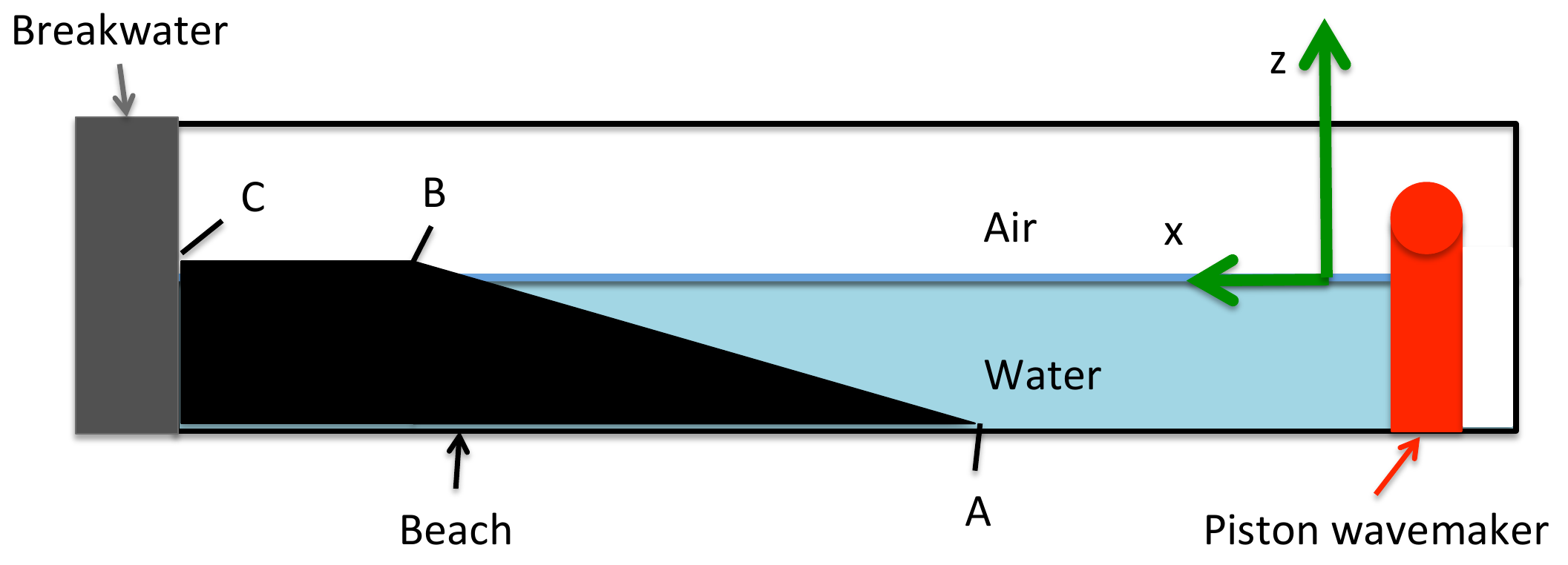}   \\ 
(b)
\end{tabular}
\end{center}
\caption{\label{nfa_domains} NFA simulations:  (a) Wedge wavemaker problem. (b) Piston wavemaker problem.}
\end{figure*}

Consider the  immiscible turbulent flow at the interface between air and water with $\rho_a$ and $\rho_w$ respectively denoting the densities of air and water.   Physical quantities are normalized by characteristic length ($L_o=D$), velocity($U_o=\sqrt{D g}$),  time ($L_o/U_o=\sqrt{D/g}$), density ($\rho_w$), and pressure ($\rho_w U_o^2$) scales.  Here, $D$ is the depth of the wave tank and $g$ is the acceleration of gravity.

Figure \ref{nfa_domains}a\&b show the computational domains for the wedge and piston wavemaker problems that are simulated using NFA.   As constructed the wave makers are entirely within the computational domains so that the fluid volumes are conserved as the wave makers move.     The flow of air and water are modeled in both simulations.   As shown in Figure \ref{nfa_domains}a, a cube is downstream of the wedge wavemaker.  The wave that is generated breaks on the cube.   As shown in Figure \ref{nfa_domains}b, a beach is located downstream of the piston wavemaker.   The wave that is generated breaks on the beach, surges up the beach, and impacts the breakwater.  The wedge wavemaker generates waves by oscillating up and down.   The piston wavemaker generates a soliton by surging forward.

Let $\alpha$ denote the fraction of fluid that is inside a cell. By definition, $\alpha=0$ for a cell that is totally filled with air, and $\alpha=1$ for a cell that is totally filled with water.   In terms of $\alpha$, the normalized density is express as
\begin{eqnarray}
\label{alpha}
\rho(\alpha) & = & \lambda + (1 - \lambda) \alpha \;\; ,
\end{eqnarray}
\noindent where $\lambda = \rho_a/\rho_w$ is the density ratio between air and water.  

Let $u_i$ denote the normalized three-dimensional velocity field as a function of normalized space ($x_i$) and normalized time ($t$).  The conservation of mass is
\begin{eqnarray}
\label{mass}
\frac{\partial \rho}{\partial t} +\frac{\partial u_j \rho}{\partial x_j} = 0 \;\; .
\end{eqnarray}

For incompressible flow,
\begin{eqnarray}
\label{density}
\frac{\partial \rho}{\partial t} +u_j \frac{\partial  \rho}{\partial x_j} = 0 \;\; .
\end{eqnarray}

Subtracting Equation (\ref{density}) from (\ref{mass}) gives a solenoidal condition for the velocity:
\begin{eqnarray}
\label{solenoidal}
\frac{\partial u_i}{\partial x_i} = 0 \;\; .
\end{eqnarray}

Substituting Equation (\ref{alpha}) into (\ref{mass}) and making use of (\ref{solenoidal}), provides an advection equation for  the volume fraction:
\begin{eqnarray}
\label{vof}
\frac{\partial \alpha}{\partial t}+ \frac{\partial}{\partial x_j} \left(u_j \alpha \right)= 0 \;\; .
\end{eqnarray}

For very high Reynolds numbers, viscous stresses are negligible, and the conservation of momentum is
\begin{eqnarray}
\label{momentum}
\frac{\partial u_i}{\partial t}+\frac{\partial}{\partial x_j} \left(u_j u_i \right)  =  -\frac{1}{\rho} \frac{\partial p}{\partial x_i}  - \frac{\delta_{i3}}{F_r^2}  \;\; ,
\end{eqnarray}
\noindent where $F_r^2 = U_o^2/(g L_o)=1$ is the Froude number.  $p$ is the pressure, and $\delta_{ij}$ is the Kronecker delta function.   

A no-flux condition is imposed on the surface of the wavemaker ($S_w$).
\begin{eqnarray}
\label{bc1}
u_i n_i = v_i n_i  \;  {\rm on} \; S_w.
\end{eqnarray}
Here, $v_i$ is the velocity of the wave maker and $n_i$ is the unit normal on the surface.   The normal velocity is zero on the walls of the wave tank ($S_o$), including the vertical walls, the side walls, the top, and the bottom.
\begin{eqnarray}
\label{bc2}
u_i n_i = 0  \;   \;  {\rm on} \; S_o .
\end{eqnarray}

The divergence of the momentum equations (\ref{momentum}) in combination with the solenoidal condition (\ref{solenoidal}) provides a Poisson equation for the dynamic pressure:
\begin{eqnarray}
\label{poisson} 
\frac{\partial}{\partial x_i} \frac{1}{\rho} \frac{\partial
p}{\partial x_i} = \Sigma \;\; ,
\end{eqnarray}
\noindent where $\Sigma$ is a source term.  The pressure is used to project the velocity onto a solenoidal field and to impose a no-flux condition on the surface of the wavemaker and the walls of the tank.  Details of the volume fraction advection, the pressure projection, and the numerical time integration are provided in \citet{dommermuth07} and \citet{dommermuth08}. Sub-grid scale stresses are modeled using an implicit model that is built into the treatment of convective terms.  The performance of the implicit SGS model is provided in \citet{nfa3}. 

\subsection{BIEM Formulation}

\begin{figure*}
\begin{center}
\begin{tabular}{c}
\includegraphics[width=0.45\linewidth]{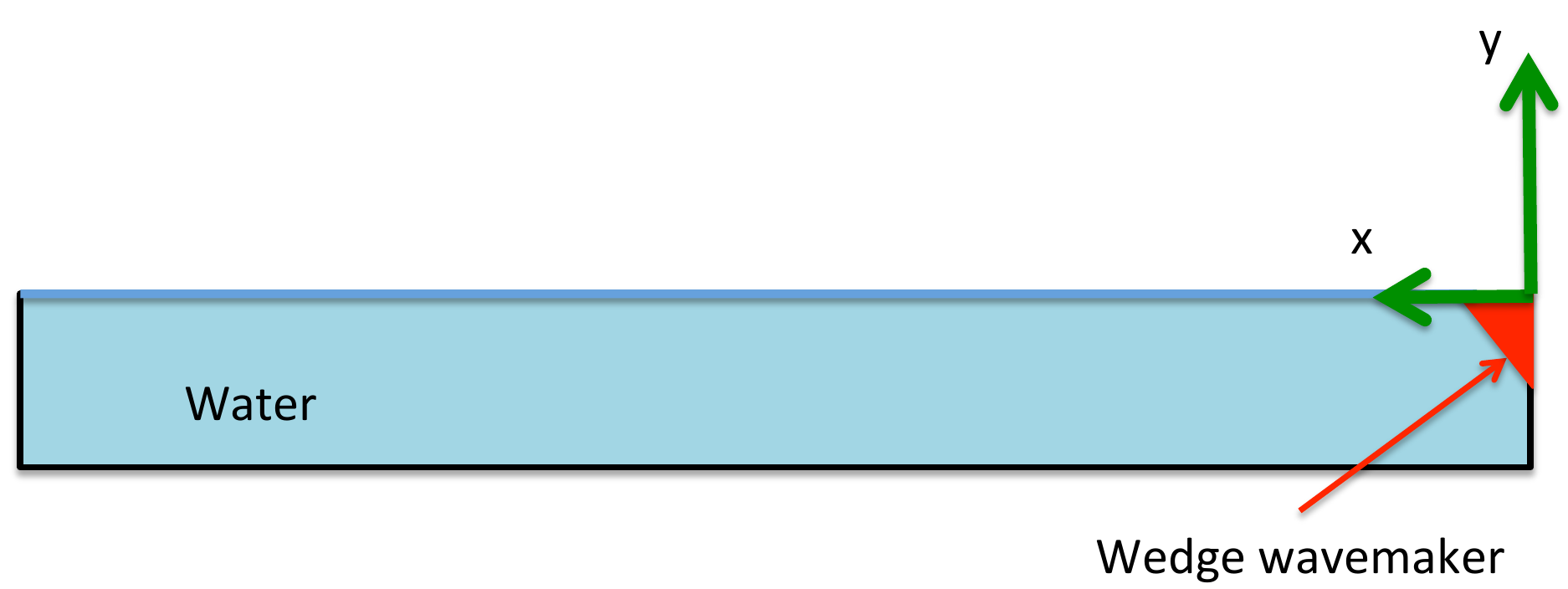}   \\
\end{tabular}
\end{center}
\caption{\label{biem_domain} BIEM wedge wavemaker problem.}
\end{figure*}

The mathematical formulation that we use is similar to \cite{vinjeandbrevig}.   Additional details are provided in \cite{dommermuth88}.  The wave tank problem is formulated as the irrotational flow in a two-dimensional tank with a wedge wavemaker at one end.     We define Cartesian coordinates with the origin at the intersection of the of the right wall with the extension of the undisturbed free-surface (y=0) into the wall where the wedge wavemaker rises up and down.   

Figure \ref{biem_domain} shows the computational domain for the wedge problem that is simulated using BIEM.   The impact of the wave on the cube is not simulated using BIEM.   NFA and BIEM solutions are compared without the cube as a check of the numerical methods.

For the BIEM solution, we define the complex potential
\begin{eqnarray}
\label{biem1}
\beta(z,t)=\phi(x,y,t)+\imath\psi(x,y,t) \; ,
\end{eqnarray}
\noindent were $z=x+\imath y$.   Since both the velocity potential $\phi$ and the stream function $\psi$ are solutions to Laplace's equation in the fluid domain, the Cauchy integral theorem can be applied to give
\begin{eqnarray}
\label{biem2}
2\pi\imath \beta(\zeta,t)=\int_C \frac{\beta(z,t)}{z-\zeta} dz \; ,
\end{eqnarray}
\noindent where the contour of integration, $C(z,t)$, is a closed path with includes the wavemaker $(S_w)$, the free-surface ($S_f$), and the walls of the tank ($S_o$).   $\zeta$ is on C.     

On the walls of the tank, the normal velocity is zero and $\psi=0$.   On the wavemaker, the normal velocity is prescribed and the stream function is given by the integral of the normal flux along the surface of the wavemaker:
\begin{eqnarray}
\label{biem3}
\psi(s)=\int^s_{s_o}  v_i n_i dl \; ,
\end{eqnarray}
\noindent where recall that $v_i$ and $n_i$ are respectively the velocity and  unit normal of the wavemaker.   $s_o$ is the lower corner of the wavemaker and $s$ runs along the wavemaker ($l$).

On the free-surface, the kinematic condition is used to track the position of the free-surface ($F(z,t)$): 
\begin{eqnarray}
\label{biem4}
\frac{d z}{d t}=\frac{\partial \beta^*}{\partial z} \; ,
\end{eqnarray}
where $d/dt \equiv \frac{\partial}{\partial t} + \nabla \phi \cdot \nabla$ is the substantial derivative and an asterisk denotes a complex conjugate.   From Bernoulli's equation, the dynamic boundary condition is zero atmospheric pressure on the free-surface:
\begin{eqnarray}
\label{biem5}
\frac{d \phi}{d t}=\frac{1}{2} \left| \frac{\partial \beta}{\partial z} \right| -y \; .
\end{eqnarray}
\noindent With the specification of the initial conditions corresponding to the fluid being at rest at $t=0$, the initial boundary-value problem for $\beta(z,t)$ and $F(z,t)$ is complete.    We solve the problem using a mixed Eulerian Lagrangian method \citep{long1976}.   At any instant of time $t$, the wavemaker's position and the stream function ($\psi(z,t)$) on the wavemaker are prescribed, and the position of the free-surface $(F(z,t))$ and the velocity potential ($\phi(z,t)$) on the free-surface are given from the integration of the kinematic (Equation \ref{biem4}) and dynamic (Equation \ref{biem5}) boundary conditions, respectively.   Equation \ref{biem2}  is used to solve for the unknown $\phi(z,t)$ on $S_w$ and $S_o$, and $\psi$ on $S_f$.   As result, we can integrate Equation \ref{biem5}  for the new Lagrangian value of $\phi(z,t+\Delta t)$ on the free-surface and the new position of the free-surface  $F(z,t+\Delta t)$ and the whole process is repeated.   Additional details of the numerical implementation are provided in \cite{dommermuth88}.

\section{Results}
\subsection{Experiments}

\begin{figure*}
\begin{center}
\begin{tabular}{cc}
(a) & (b)\\
\includegraphics[width=0.45\linewidth]{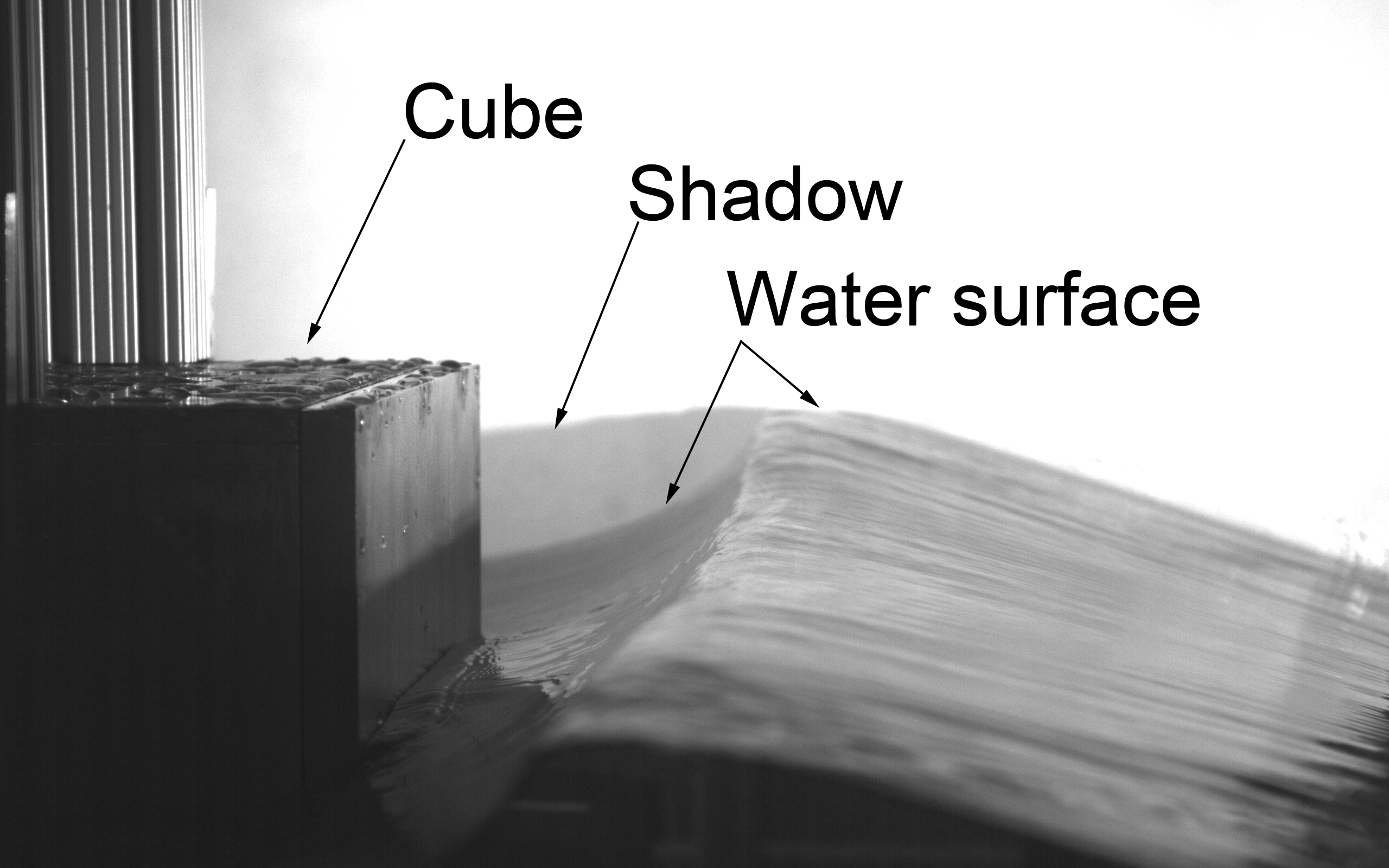}&\includegraphics[width=0.45\linewidth]{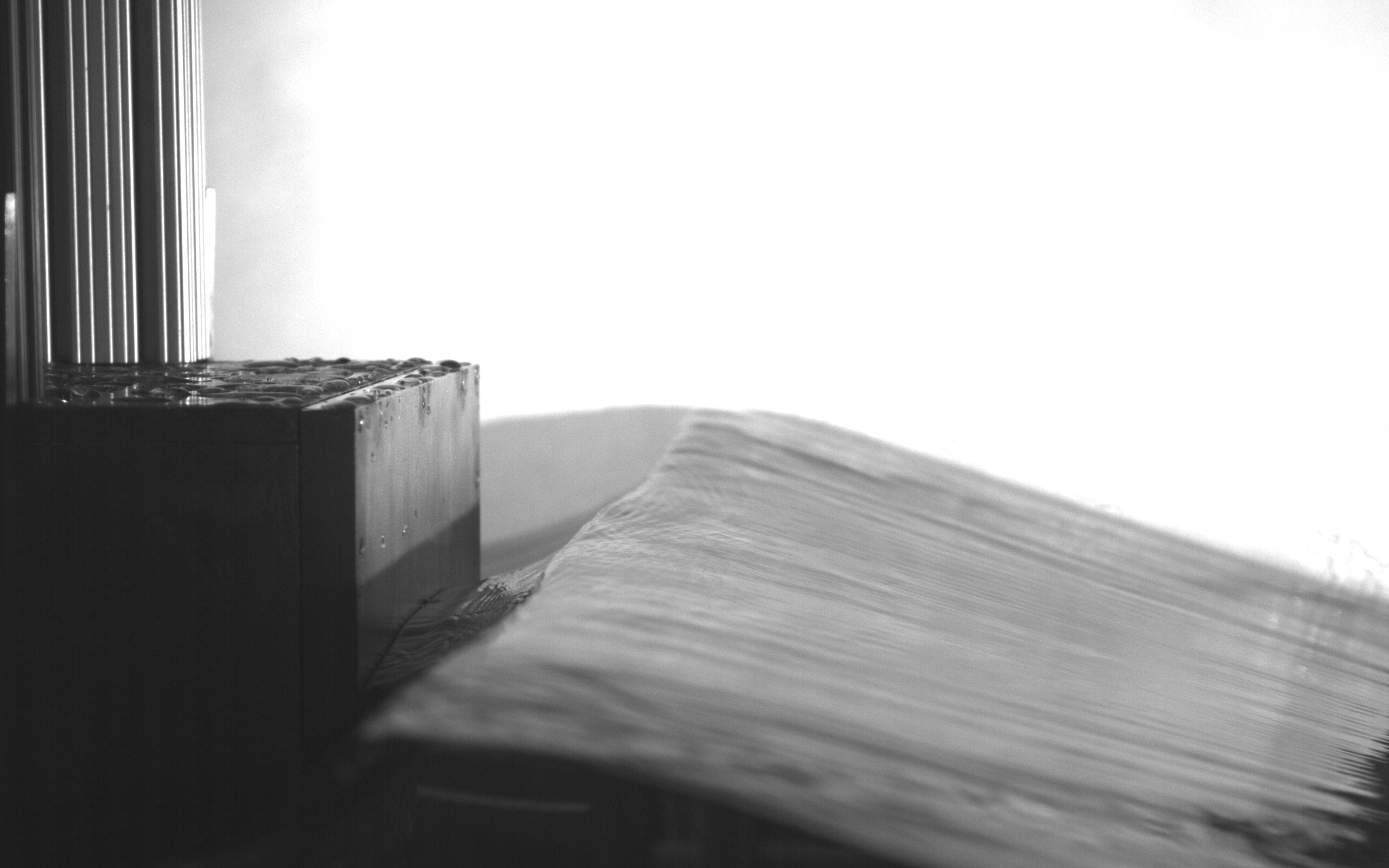}\\
(c) & (d)\\
\includegraphics[width=0.45\linewidth]{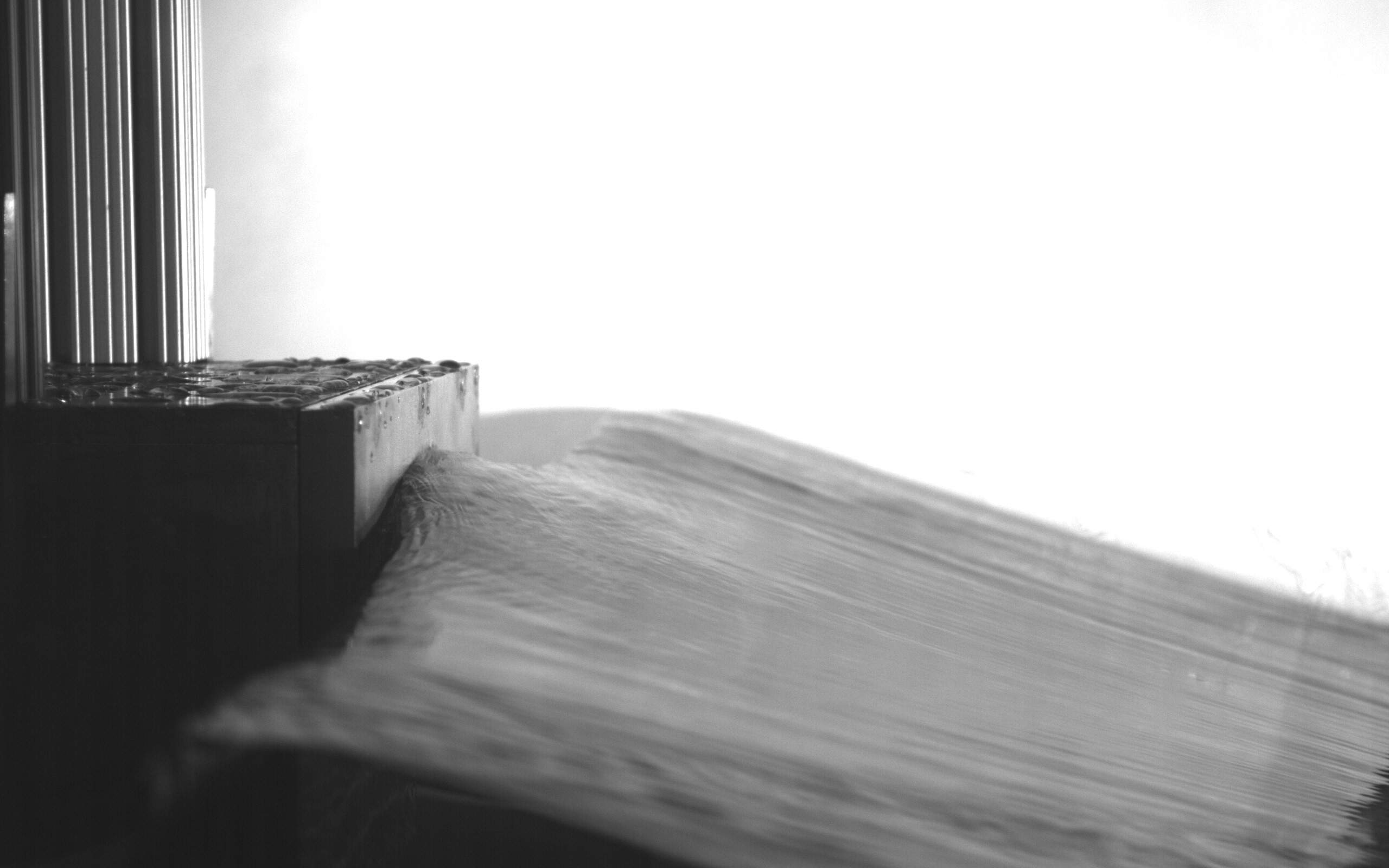}&\includegraphics[width=0.45\linewidth]{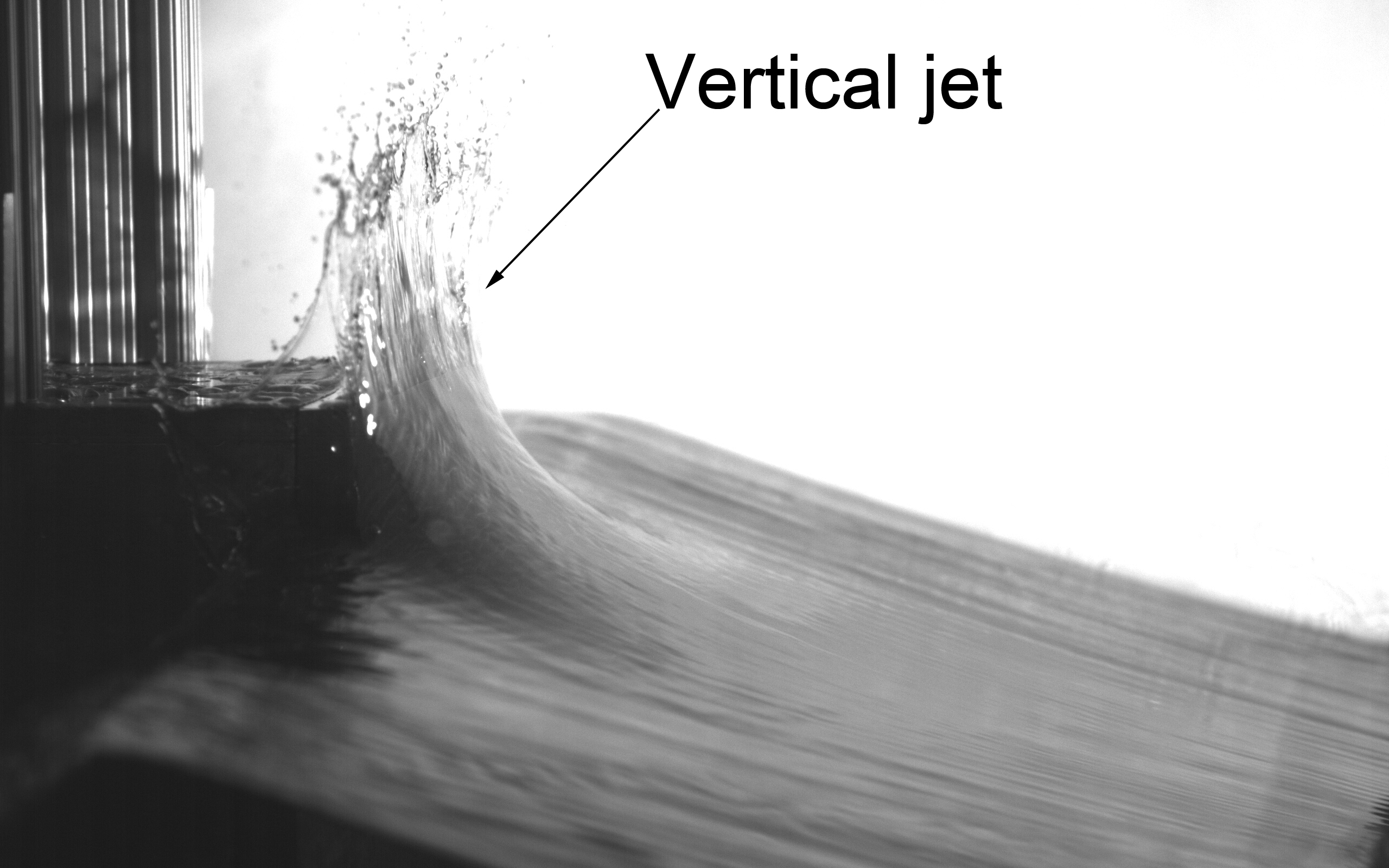}
\end{tabular}
\end{center}
\vspace*{-0.2in} \caption{A sequence of four images from a high-speed movie of the impact of a breaking wave on the cube, $X_{\mbox{cube}} =
  647$~cm, $Z_{\mbox{cube}}=0$~cm.  In this
  case, the plunging jet forms as it hits the front face of the cube,
  thereby, creating a fast-moving vertical jet.  The time interval
  between the photographs in 33.333~ms.} \label{fig:photos}
\end{figure*}

\begin{figure*}
\begin{center}
\begin{tabular}{c}
(a) \\
\includegraphics[width=0.9\linewidth,trim= 1.3in 3.0in 1.0 3.5in,clip=true]{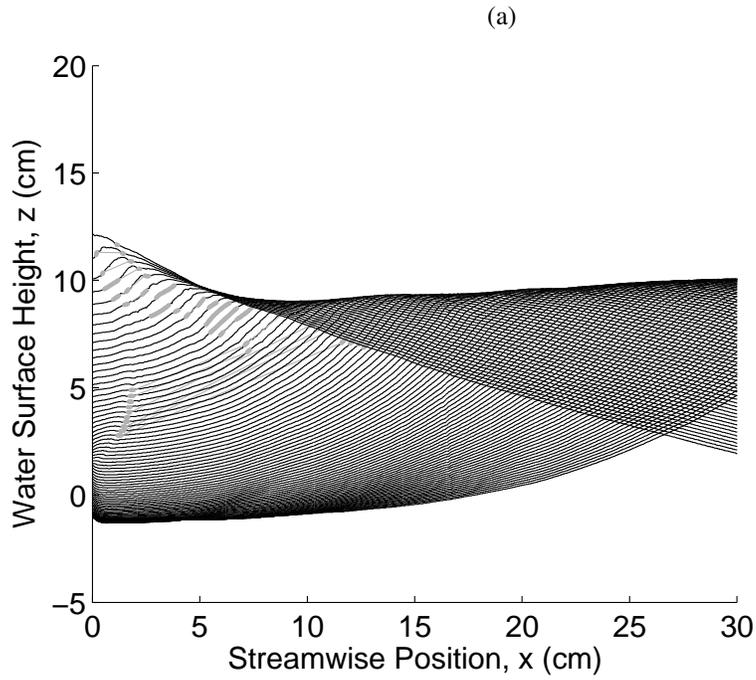}\\
(b)\\
\includegraphics[width=0.9\linewidth,trim= 1.3in  3.25in 1.0 3.5in,clip=true]{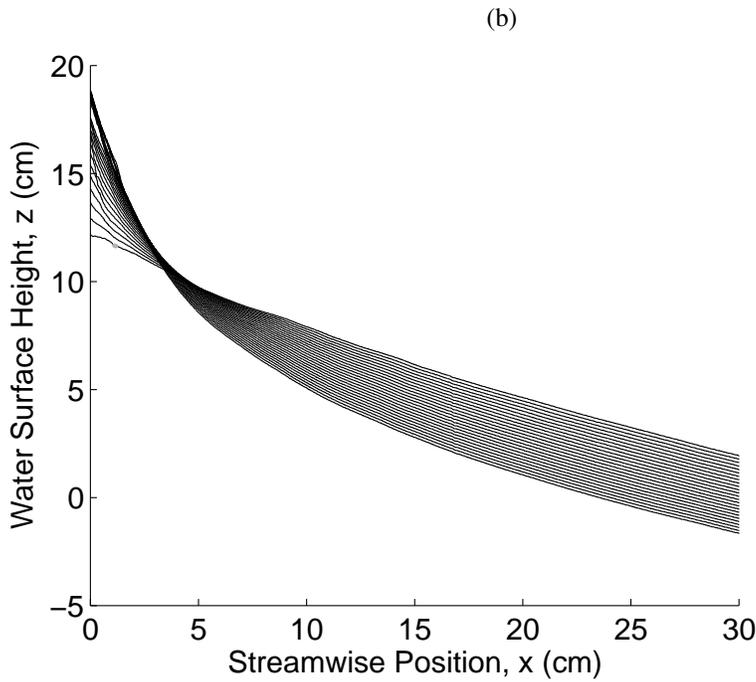}
\end{tabular}
\end{center}
\vspace*{-0.2in} \caption{Sequence of profiles of the water-surface
  during impact of the wave, $X_{\mbox{cube}} = 647$~cm,
  $Z_{\mbox{cube}}=0$~cm.  The time between profiles is 3.333~ms.  For
clarity, the earlier profiles, up to the point of formation of the
vertical jet, are shown in (a) and the later profiles, after the point
of formation of the vertical jet, are shown in (b).  The small gray
areas on each curve are locations where the profile had to be created
by spline fitting because the image of the intersection of the light sheet and the
water surface was obscured by part of the wave crest.} \label{fig:profiles}
\end{figure*} 

\begin{figure*}
\begin{center}
\hspace*{-0.0in}\includegraphics[width=1.3\linewidth,trim= 1.3in 3.0in 1.0 3.0in,clip=true]{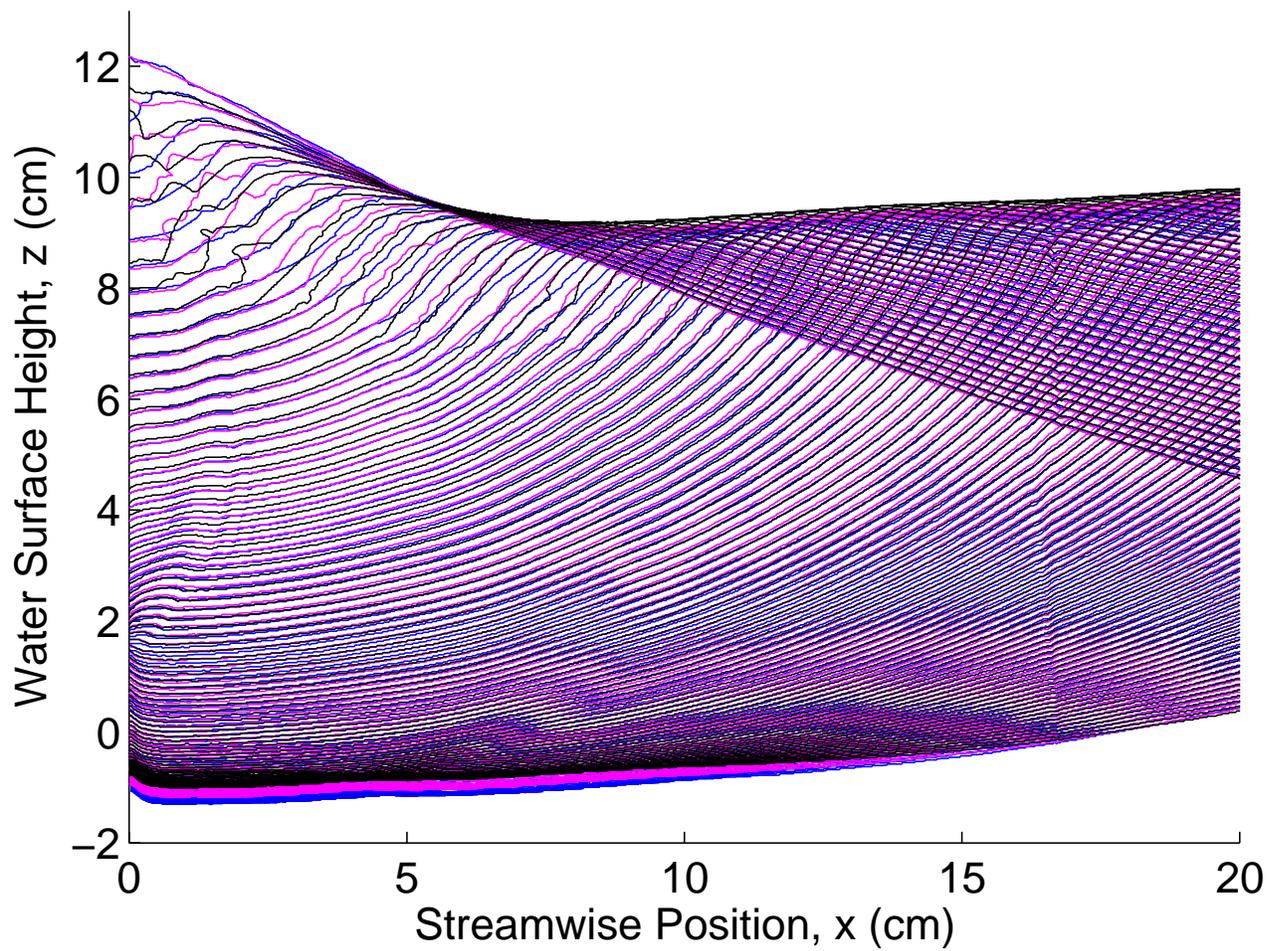}\\
\end{center}
\vspace*{-0.2in} \caption{Sequence of profiles of the water-surface
  during impact of the wave up to the point of formation of the
  vertical jet, $X_{\mbox{cube}} = 647$~cm,
  $Z_{\mbox{cube}}=0$~cm.  The time between profiles is 3.333~ms.
  Profiles are shown for three experimental runs.} \label{fig:ThreeRuns}
\end{figure*}

\begin{figure*}
\begin{center}
\begin{tabular}{cc}
(a) & (b) \\
\includegraphics[width=0.55\linewidth,trim= 1.1in 3.0in 1.0 3.25in,clip=true]{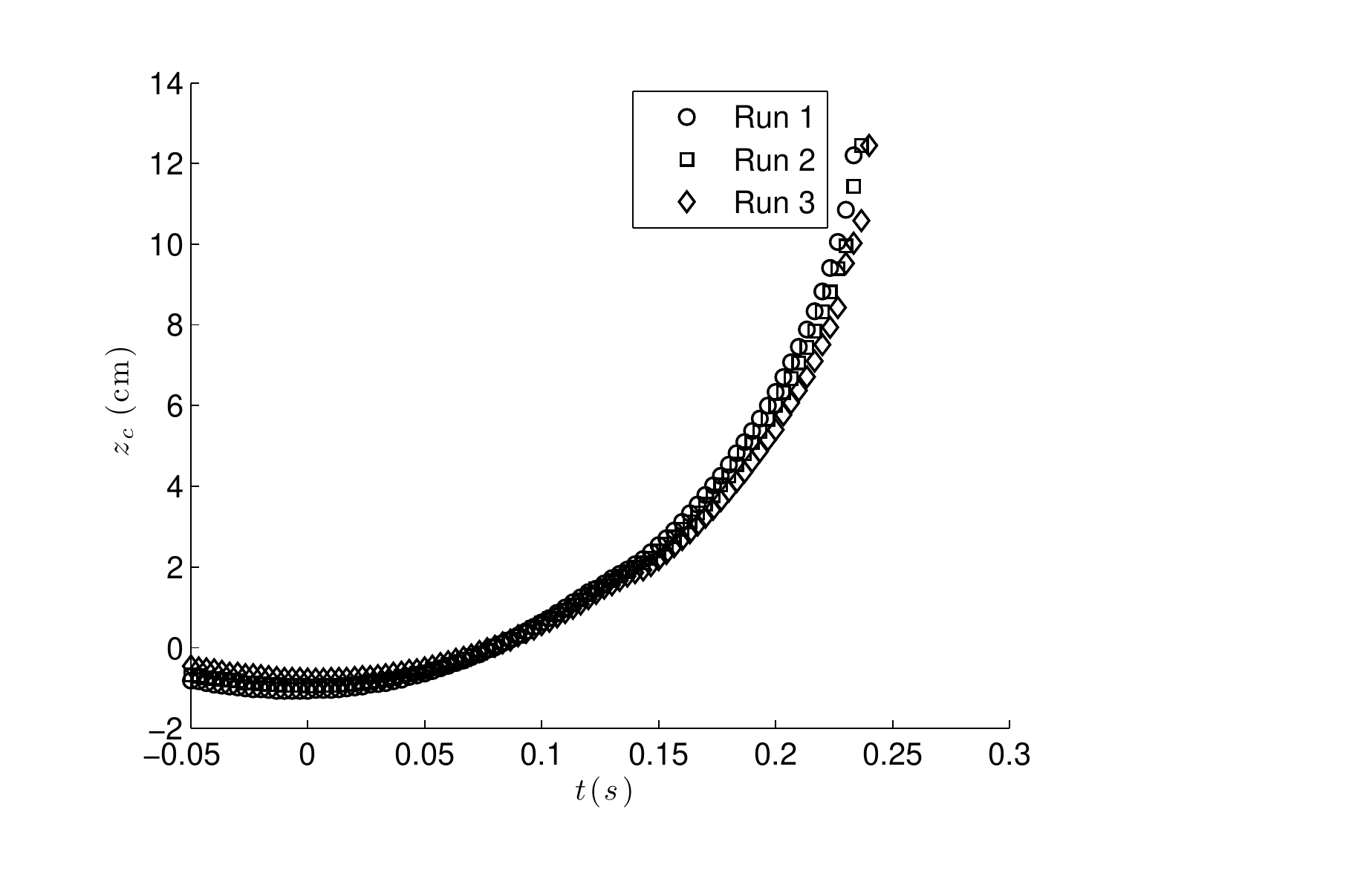}&
\includegraphics[width=0.55\linewidth,trim= 1.1in  3.0in 1.0 3.25in,clip=true]{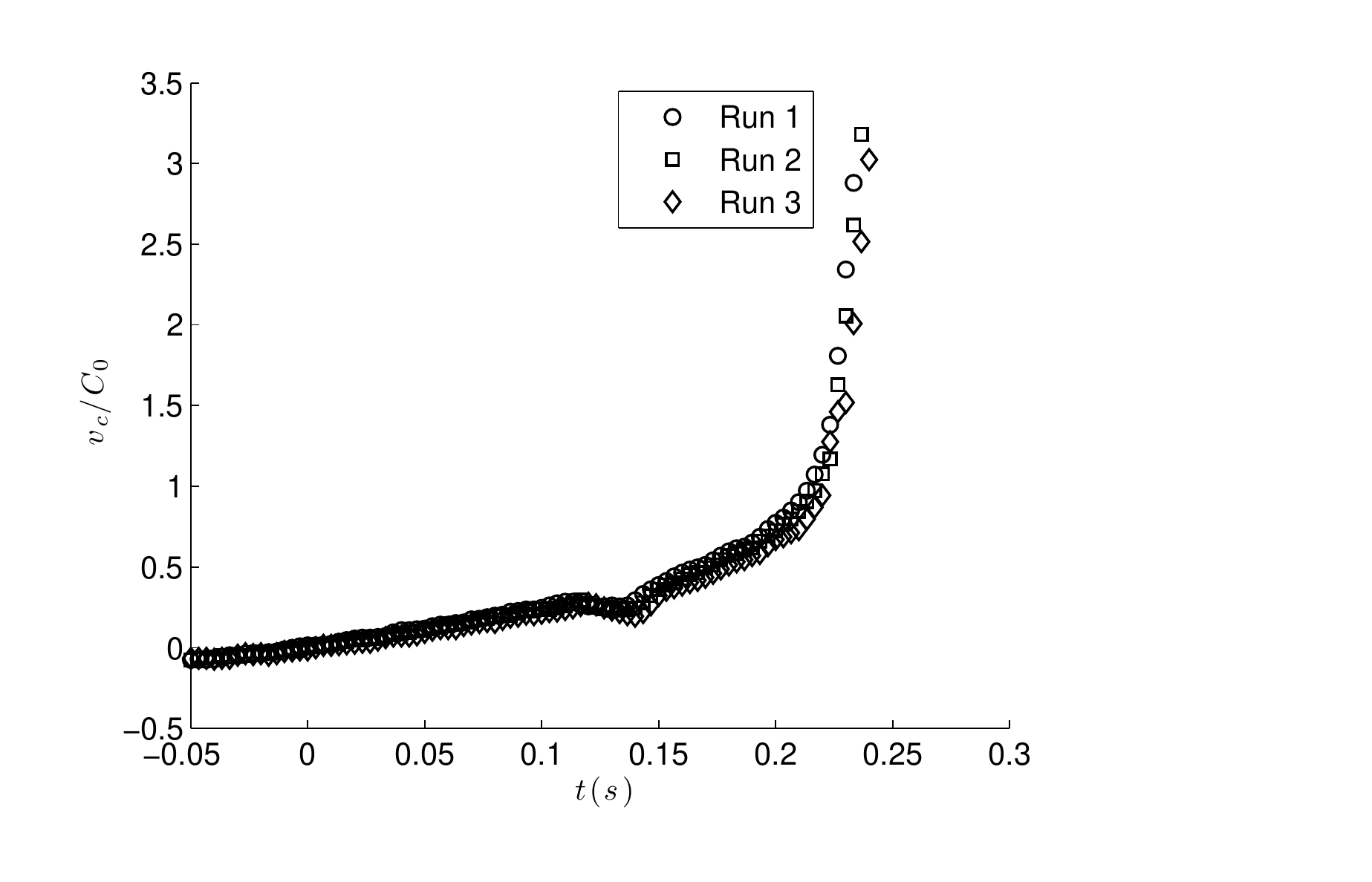}
\end{tabular}
\end{center}
\vspace*{-0.2in} \caption{Contact-point behavior.  (a). Contact-point
  height, $z_c$, versus time.  (b) Contact-point vertical velocity versus time.  Data is shown for three runs
  to demonstrate the repeatability of the contact-point motion.
  $X_{\mbox{cube}} =647$~cm and   $Z_{\mbox{cube}}=0$~cm.} \label{fig:zcvc}
\end{figure*} 

A sequence of four images from a white-light movie of the impact of the
wave with the front face of the cube located at 647~cm from the back face of the wave
maker is shown in Figure~\ref{fig:photos}.  In calm water, the horizontal mid  plane of the cube is
coincident with the water free-surface.  As the wave approaches the
front face of the cube, the region between the wave crest and the
cube surface forms an arc with upward curvature.  As the wave progresses
toward the cube face, this arc closes and a jet forms with vertical
velocity, parallel to the cube face.

The LIF movies for this case were processed and the resulting sequence
of the surface profile shape from the contact-point on the front face of the
cube to a region upstream of the wave crest is shown in
Figure~\ref{fig:profiles}(a) and (b).  The profile shapes  up to the moment that the vertical jet
is formed are given in Figure~\ref{fig:profiles}(a) and the subsequent
profiles showing the jet are given in Figure~\ref{fig:profiles}(b).

An important issue for comparison with numerical calculations is the
repeatability of this surface profile history.  To examine this, the wave
impact for the above condition was repeated three times.  Profile
histories from these three runs up to the time of formation of the vertical jet are shown in
Figure~\ref{fig:ThreeRuns}.  As can be seen in the figure, the
profiles are very close, except at the forward face of the crest where
the amount of breaking before impact is seen to vary a bit from run to
run.  This effect is likely due to residual water motions in the wave
tank.  The run depicted in Figures~\ref{fig:photos} and \ref{fig:profiles} was the
first run in the series.  Once the first run was completed, the tank was filtered and
skimmed for about 20 minutes and then allowed to calm for about 10
minutes before the next run.  In future experiments, this calming period will be increased
and variations in breaking from run to run are likely to decrease.

The height ($z_c$) and vertical velocity ($v_c$) of the contact-point of the water
free-surface on the front face of the cube are plotted versus time in
Figures~\ref{fig:zcvc}(a) and (b), respectively.  The water-surface
height rises continuously during the impact process.  The velocity of
the contact-point rises slowly at first but after reaching about 1~m/s
it rises nearly linearly to about 3~m/s in about 0.02~s, yielding an average
vertical acceleration on the order of 100~m/s$^2$, about $10g$.

\subsection{Numerical Simulations}
\subsubsection{Wave Impacting a Cube}

NFA predictions of a three-dimensional breaking wave impacting on a cube are compared to experimental measurements.   NFA results of a two-dimensional run without the cube present are also compared to BIEM and a set of experiments that had been performed without the cube present.   The tank depth is 0.91 m, which is used to normalize the NFA and BIEM simulations.    The cube is half immersed. 

\begin{figure}[h]
\begin{center}
\begin{tabular}{c}
\includegraphics[width=1\linewidth]{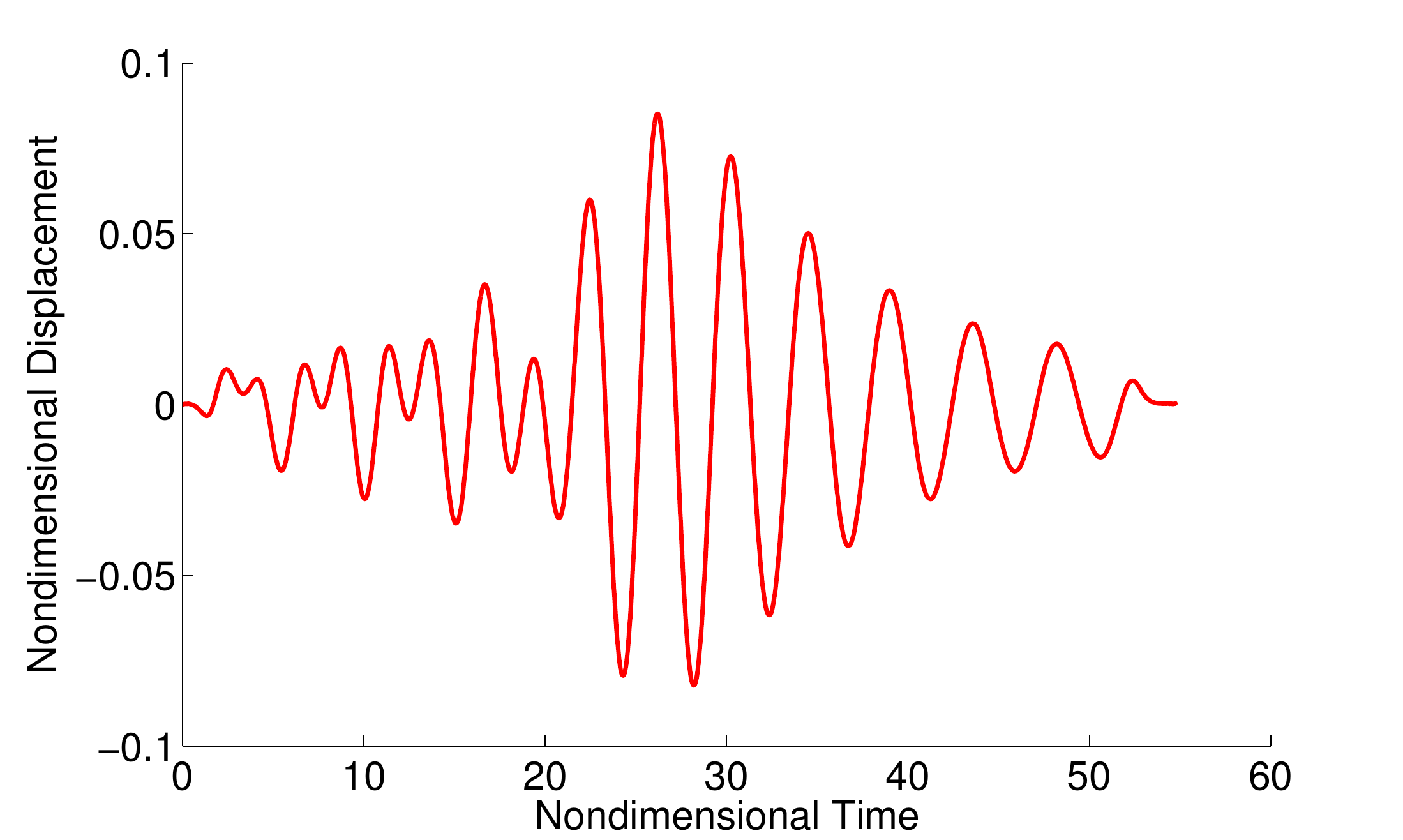}   \\
\end{tabular}
\end{center}
\caption{\label{wedge_motion} Wedge wavemaker motion.}
\end{figure}

The computational domain for the NFA simulation of the wedge-wavemaker problem is shown in Figure \ref{nfa_domains}a.  The initial depth of the tip of the wedge wavemaker is 0.4887 in non-dimensional units.    Figure \ref{wedge_motion} shows the vertical motion of the tip of the wedge wavemaker in non-dimensional units.  The non-dimensional length, width, depth of water, and height of air of the NFA computational domain are respectively 11.92, 1.2418, 1, and 0.75.  The smallest grid spacing ($\Delta x = 0.0041$), along the $x-$axis is near the wavemaker at $x=0$ and near the front face of the cube at $x=7.11$.    The grid spacing is clustered along the y-axis from $-0.2 \leq y \leq 0.2$ such that $\Delta y = 0.0041$ in this region.   The smallest grid spacing ($\Delta z = 0.0041$) along the $z-$axis is near the mean waterline ($z=0$).    The non-dimensional lengths of the edges of the cube are 0.335.  The number of subdomains along the $x$, $y$, and $z-$axes are respectively 24, 6, and 4 for the three-dimensional case and 24, 1, and 4 for the two-dimensional case.  Each subdomain has $64\times32\times64$ grid points.   The total number of grid points is 75,497,472 for the three-dimensional run and 393,216 for the two-dimensional run.    The time step is 0.001.  The simulation is run for 60,001 time steps. The density ratio of air to water is $\lambda=0.001207$.    The 3D simulation takes 72 wall-clock hours.  576 processors on the Cray XT4  at the U.S. Army Engineering Research and Development Center (ERDC) supercomputing center are used to perform the simulation. The 2D case was run on an intel desktop and took 3 hours of wall clock time using 6 cores.

The computational domain for the BIEM simulation of the wedge-wavemaker problem is shown in Figure \ref{biem_domain}.  For the BIEM simulation, the length and depth of the tank are respectively 11.92, 1.2418, and 1.   The BIEM simulation uses 700, 10, 125, 20, and 40 panels on respectively the free-surface, the far wall, the bottom, the portion of the wall that is beneath the wavemaker, and on the wavemaker.   The time step of the BIEM solution is 0.01.  The BIEM solution is run for 8,000 time steps.  Five-point smoothing of the velocity potential and the position of the free-surface is applied every 5 time steps.   The free-surface is regridded every 5 time steps. 

\begin{figure}[h!t]
\begin{center}
\begin{tabular}{rc}
\raisebox{1.0cm}{(a)} & \includegraphics[trim = 12mm 8mm 5mm 0mm, width=0.9\linewidth]{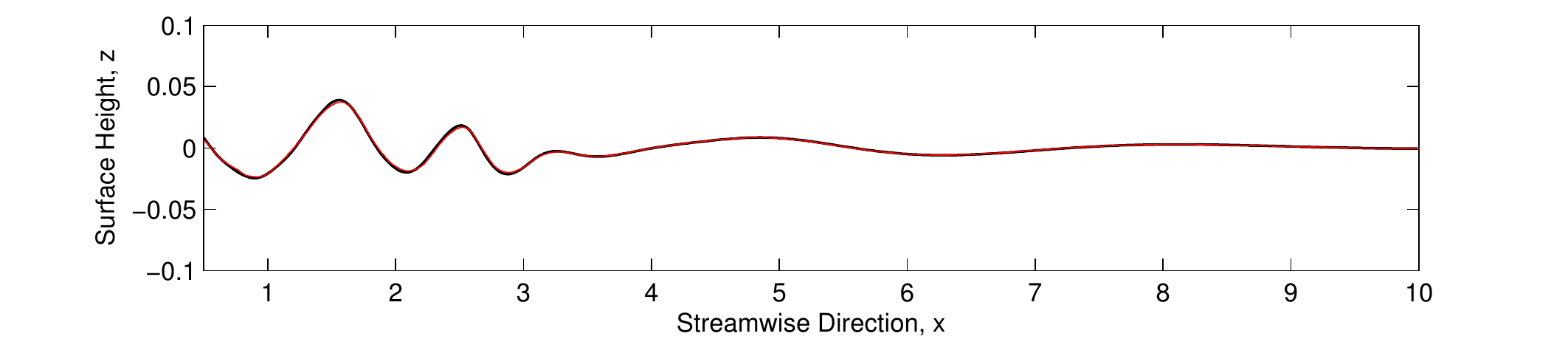}   \\
\raisebox{1.0cm}{(b)} & \includegraphics[trim = 12mm 8mm 5mm 0mm,width=0.9\linewidth]{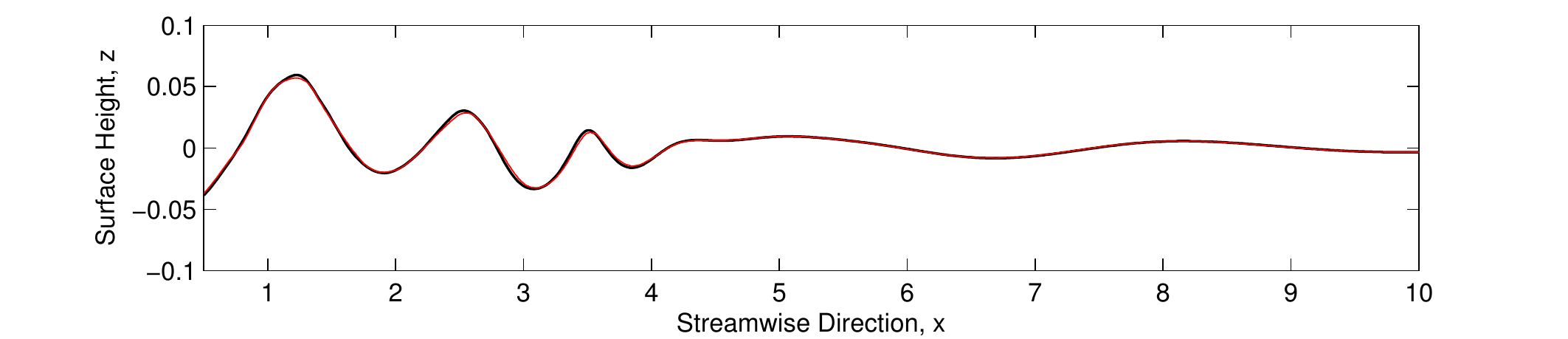}   \\
\raisebox{1.0cm}{(c)} & \includegraphics[trim = 12mm 8mm 5mm 0mm,width=0.9\linewidth]{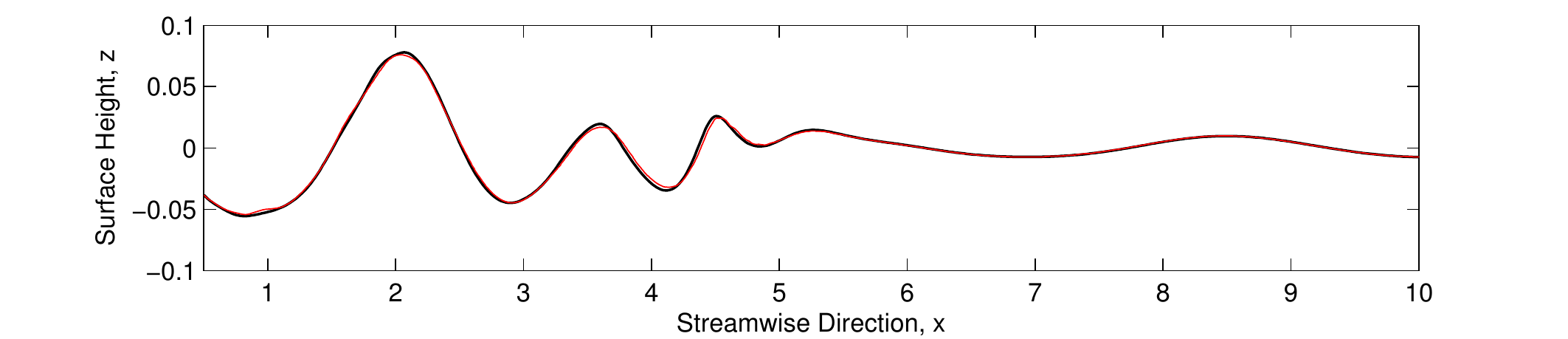}   \\
\raisebox{1.0cm}{(d)} & \includegraphics[trim = 12mm 8mm 5mm 0mm,width=0.9\linewidth]{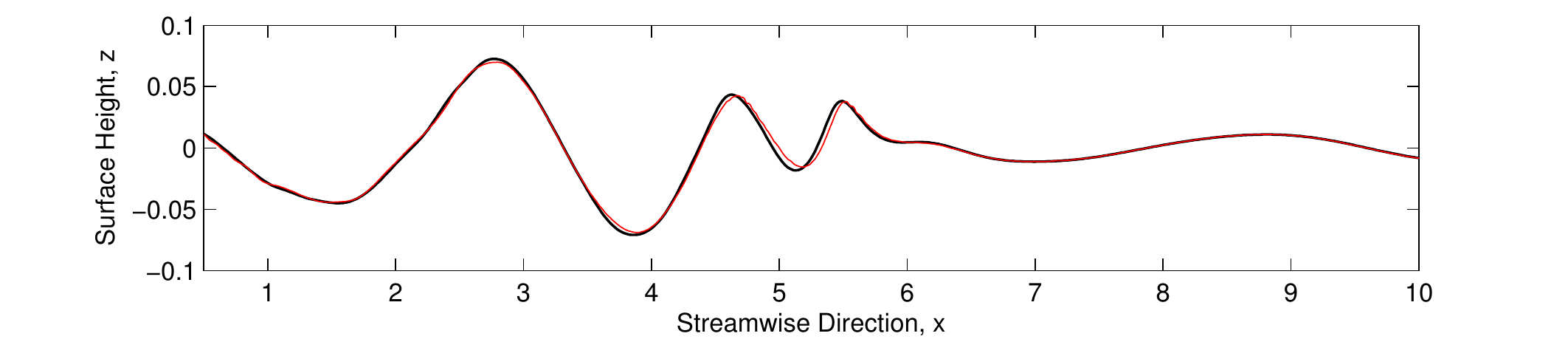}   \\
\raisebox{1.0cm}{(e)} & \includegraphics[trim = 12mm 8mm 5mm 0mm,width=0.9\linewidth]{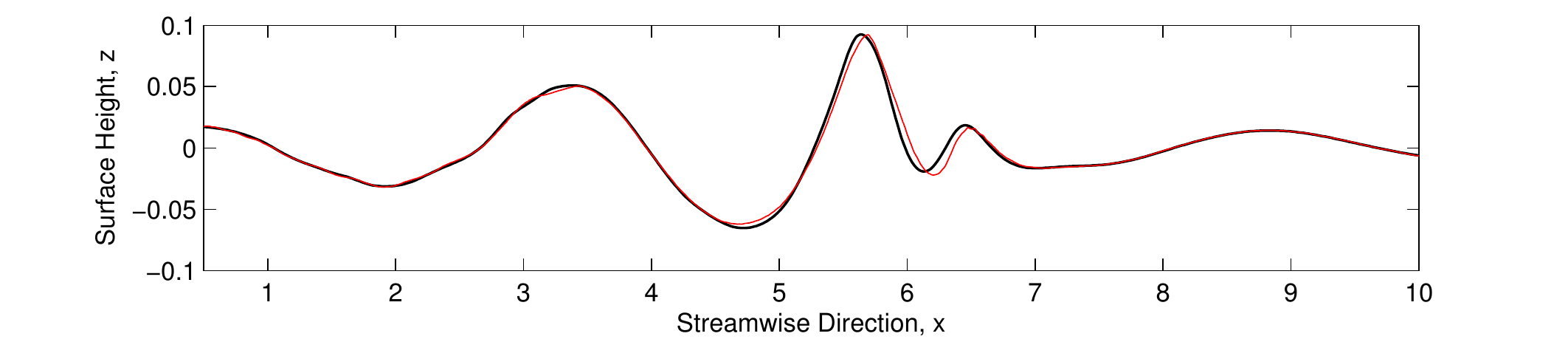}   \\
\raisebox{1.0cm}{(f)}  & \includegraphics[trim = 12mm 0mm 5mm 0mm,width=0.9\linewidth]{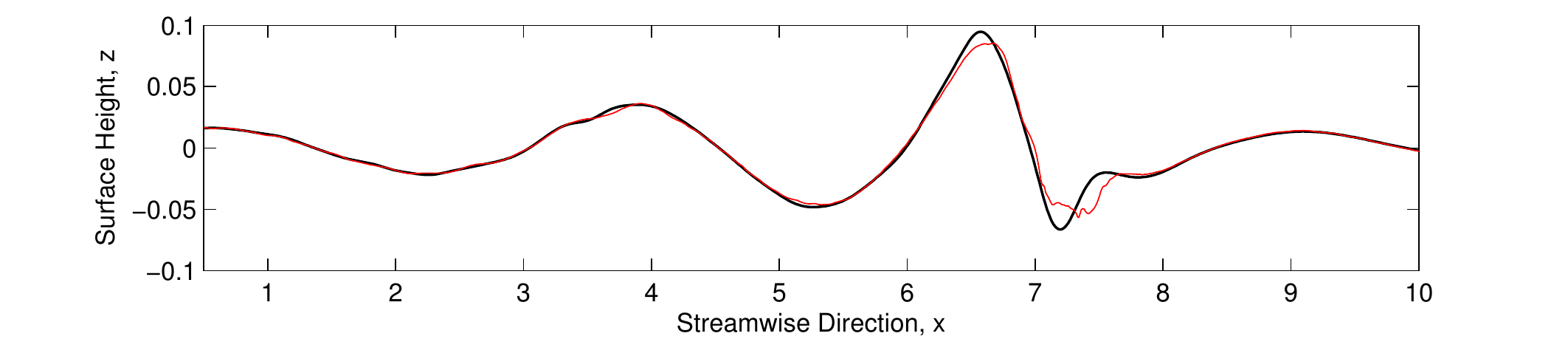}
\end{tabular}
\end{center}
\caption{\label{nfa_vs_biem} NFA (red lines) versus BIEM (black lines).   (a) $t=20$, (b) $t=25$, © $t=30$, (d) $t=35$, (e) $t=40$, and (f) $t=45$. }
\end{figure}

Figure \ref{nfa_vs_biem} compares NFA and BIEM predictions of the free-surface elevation.  There is good agreement between the two methods.   The NFA results show some evidence of wave breaking in Figure \ref{nfa_vs_biem}f.

\begin{figure}[h!t]
\begin{center}
\begin{tabular}{rc}
\raisebox{1.0cm}{(a)} & \includegraphics[trim = 50mm 0mm 0mm 4mm,clip=true,width=0.9\linewidth]{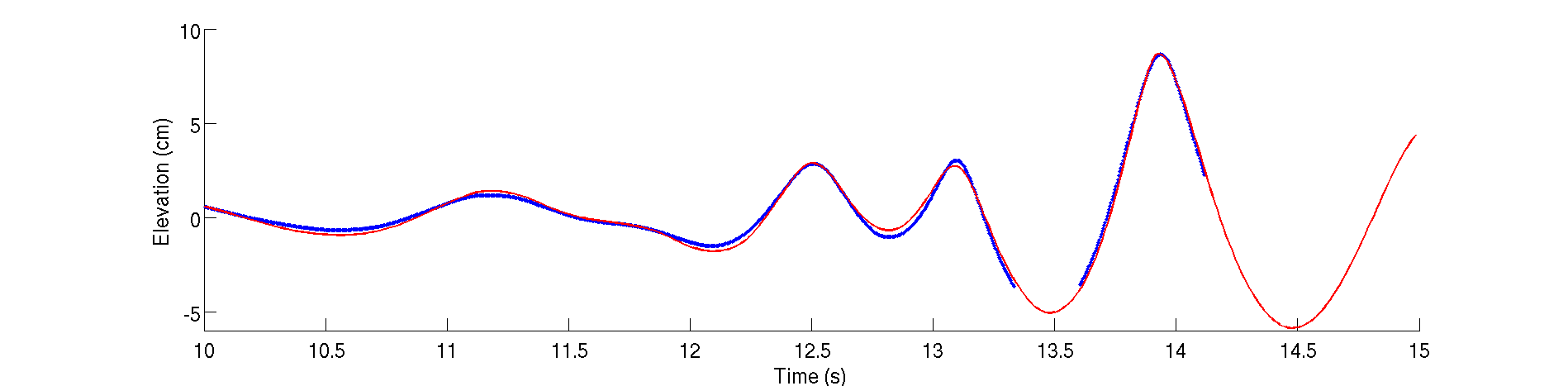}   \\
\raisebox{1.0cm}{(b)} & \includegraphics[trim = 50mm 0mm 0mm 4mm,clip=true,width=0.9\linewidth]{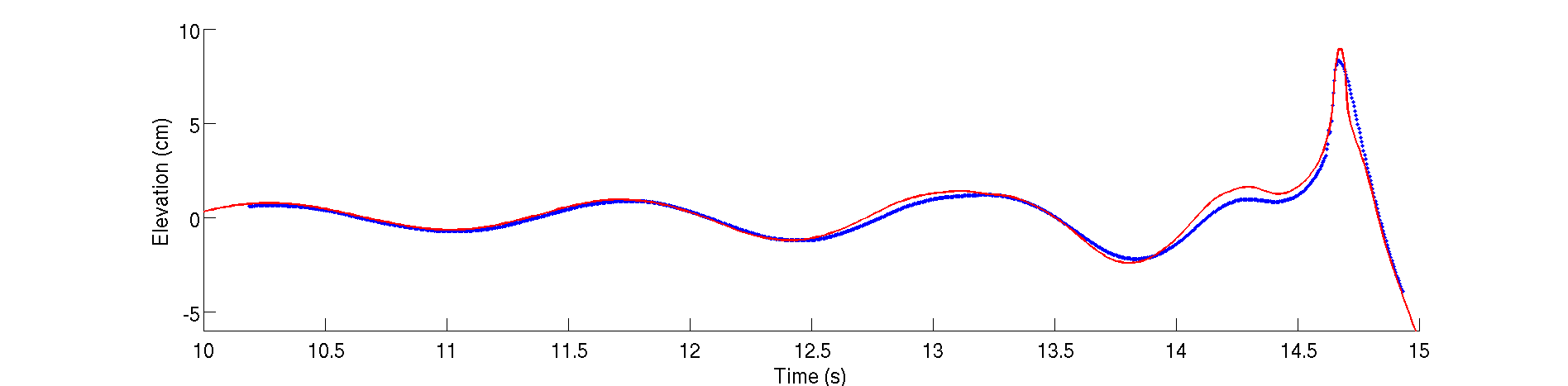}   \\
\end{tabular}
\end{center}
\caption{\label{nfa_vs_waveprobe} NFA (red lines) versus Experiments (blue lines).   (a) $x=561.75$ cm, (b) $x=670.5$ cm}
\end{figure}

To analyze the ability of NFA to capture the various stages of the converging wave packet, cameras were mounted at two different positions away from the wavemaker. The video from these cameras was analyzed in much the same way as the video from the impact focused camera, except they were focused on obtaining a single position over a longer time period. The result is wave probe like measurements during the wave packet convergence at two different distances from the wavemaker.  Figure \ref{nfa_vs_waveprobe} shows the free-surface height plotted against time for 561.75 and 670.5 cm from the back of the wavemaker. NFA is represented by the red line and the experimental height is shown as the blue line. The break in the blue line corresponds to when the free-surface left the field of view of the camera. This data was collected without the cube present, which allowed for NFA to be run with only two-dimensions. NFA shows good agreement with the experimentally measured wave height. 

\begin{figure*}[h!]
\begin{center}
\includegraphics[trim = 0mm 0mm 0mm 0mm,clip=true,width=1\linewidth]{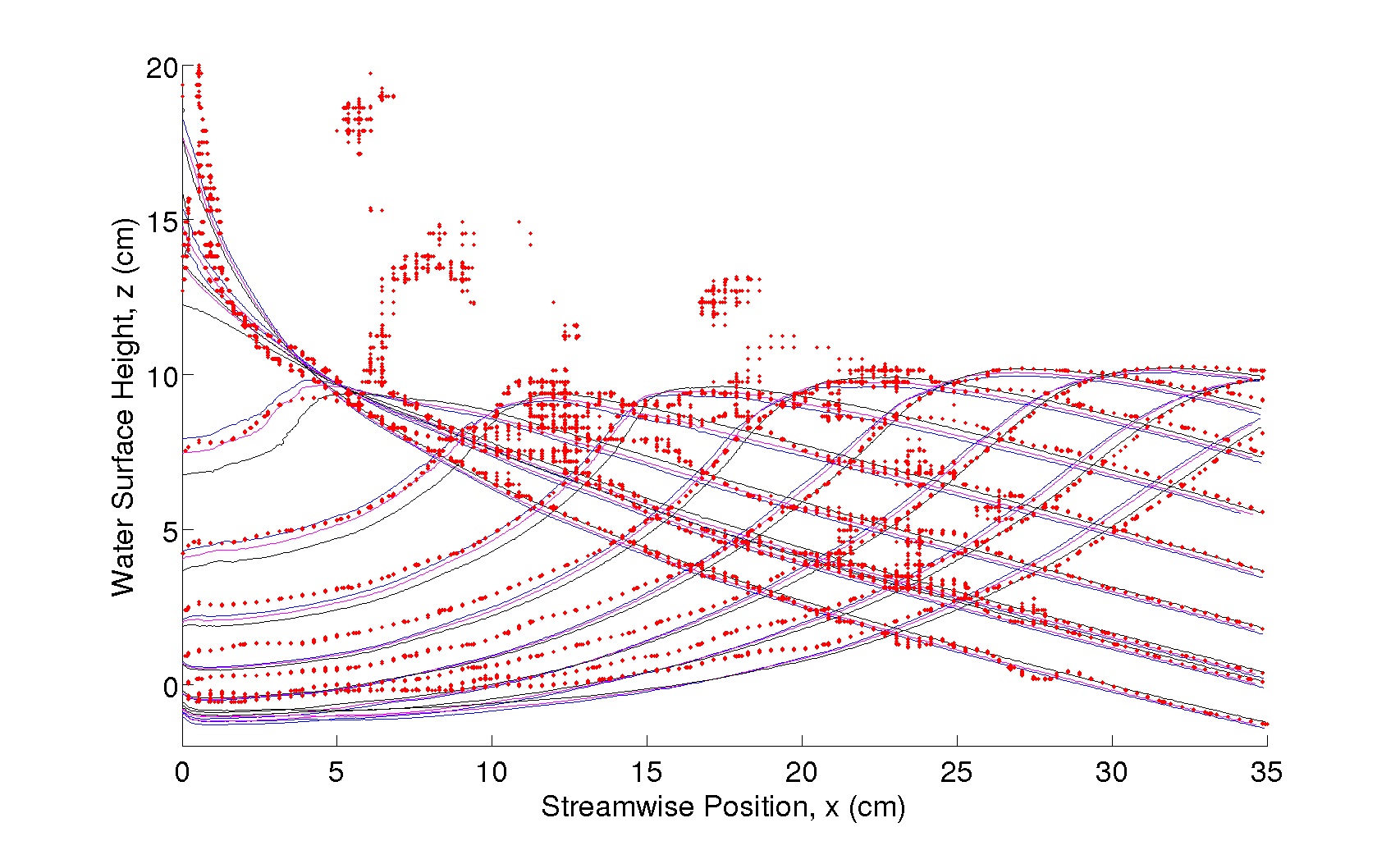}   
\end{center}
\vspace*{-0.3in}\caption{\label{nfa_vs_profiles} NFA (red dots) versus Experiments (blue, magenta and black lines).}
\end{figure*}

The three-dimensional NFA simulation of the wave impact on the cube is compared to the experimentally obtained wave profiles in Figure \ref{nfa_vs_profiles}. Select wave profiles from Figure \ref{fig:ThreeRuns} were chosen to match up with the specific NFA time steps that were periodically output.  The NFA prediction is represented by red dots in this figure. The NFA simulation compares well to the experimentally measured profiles. The trough of the wave before impact is slightly higher in the NFA simulation, but this quantity is also shown to vary between different experimental runs.

The complex motion of the wedge wavemaker contained a wide range of frequency content used to create the converging wave packet. It was necessary to turn off a key smoothing routine in NFA to preserve the entire range of frequency content due to the low resolution of the wave crest. This density-weighted velocity smoothing algorithm \cite{nfa1} is typically used to deal with discontinuity in the tangential velocity across the air-water interface. It prevents the air velocities from tearing at the free surface, and was left off for these simulations. As a result the wave breaks up earlier than is seen in the experiments. The red dots above the wave profiles in Figure \ref{nfa_vs_profiles} are the spray that has torn off of the crest of the breaking wave.

%\clearpage

\subsubsection{Wave Impacting a Breakwater}

As an illustration of NFA's ability to predict wave propagation and wave-impact loading using wave maker motion as input, NFA predictions are compared to the Oregon State University (OSU) O.H. Hinsdale Wave Research Laboratories Tsunami experiments \citep{Oshnack09}.  In the experiments, a piston-type wavemaker generates a soliton that
propagates down a wave flume, runs up a small beach, and impacts with a breakwater (see Figure \ref{nfa_domains}b). The soliton is 1.2~m high in a water depth of 2.29~m and travels over 61~m before hitting the breakwater.  The slope of the beach is 1 to 12. 

\begin{figure}[h!t]
\begin{center}
\includegraphics[width=1\linewidth,trim= 0cm 0cm 0cm .5cm,clip=true]{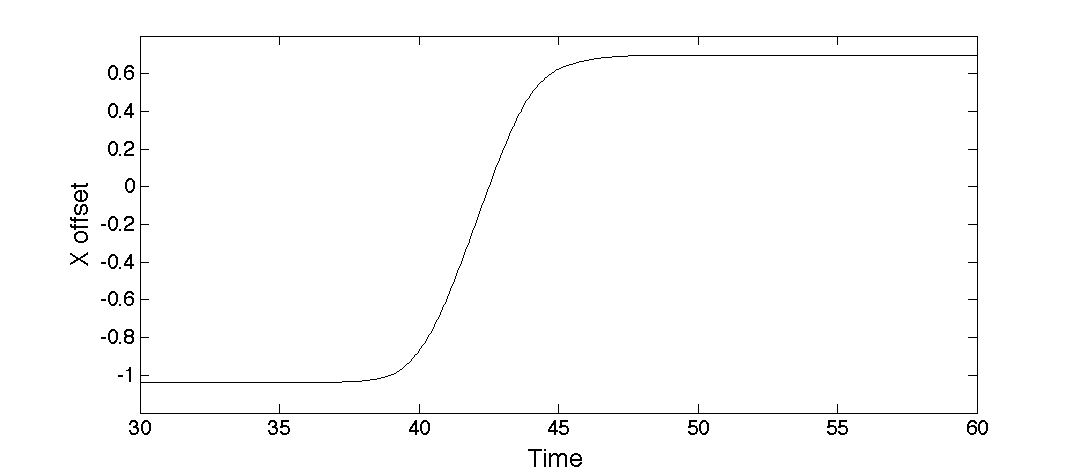}   \\
\end{center}
\vspace*{-0.2in}\caption{\label{piston_motion} Piston wavemaker motion.}
\end{figure}

In normalized coordinates, points (A), (B), and (C) in Figure \ref{nfa_domains}b are respectively located at (x,z)=(11.301,\mbox{-1.0}), (23.668,0.030568), and (26.738, 0.030568). Figure \ref{piston_motion} shows the horizontal motion of the front face of the wavemaker in non-dimensional units. The motion of the wavemaker does not begin until after $t=36$, which is the time offset when the NFA simulation starts. As the wavemaker surges forward, a large volume of water is pushed ahead of the wavemaker to form a soliton.

\begin{figure*}[h!t]
\begin{center}
\begin{tabular}{ccc}
(a) & (b) & (c)  \\
\includegraphics[width=0.3\linewidth]{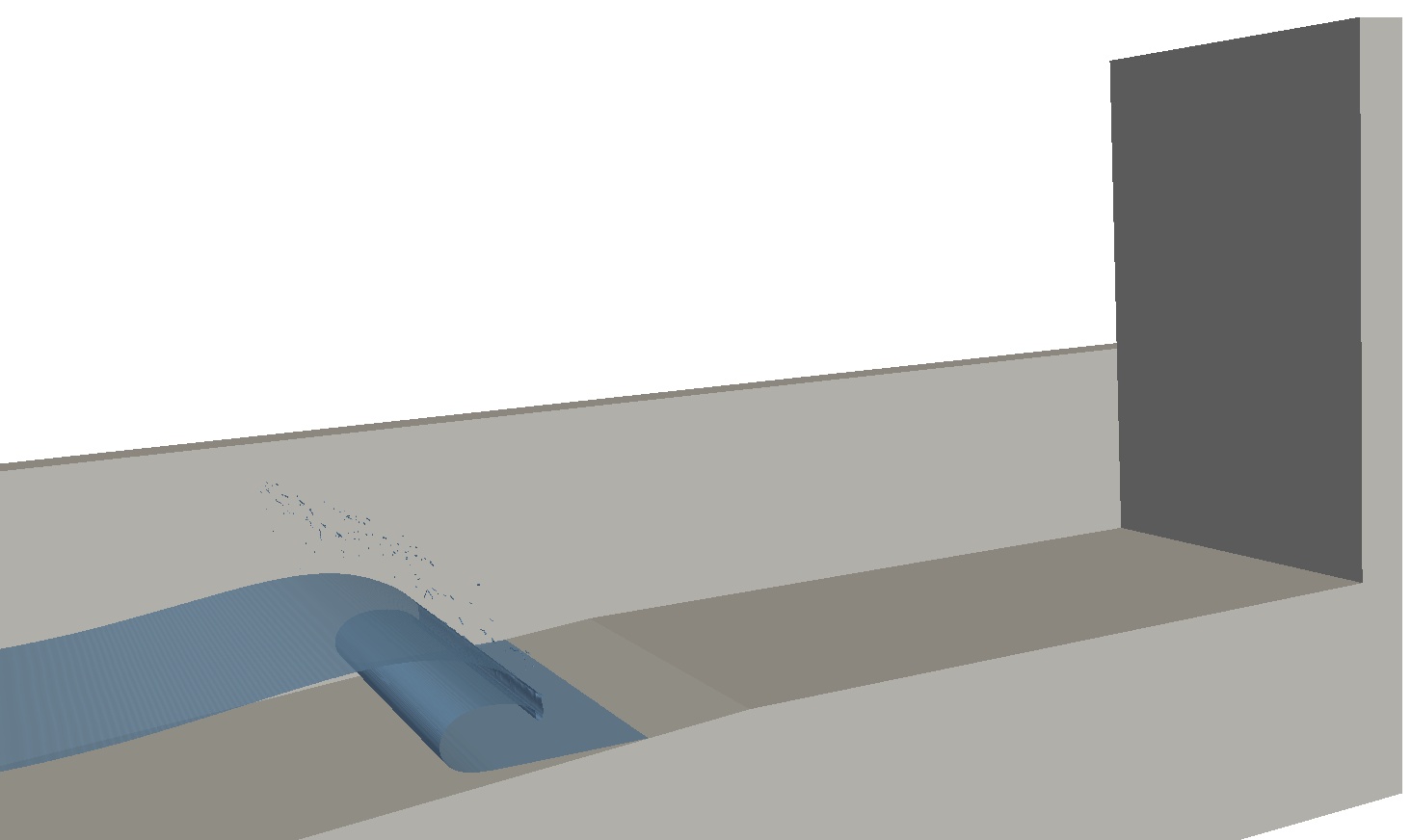}&\includegraphics[width=0.3\linewidth]{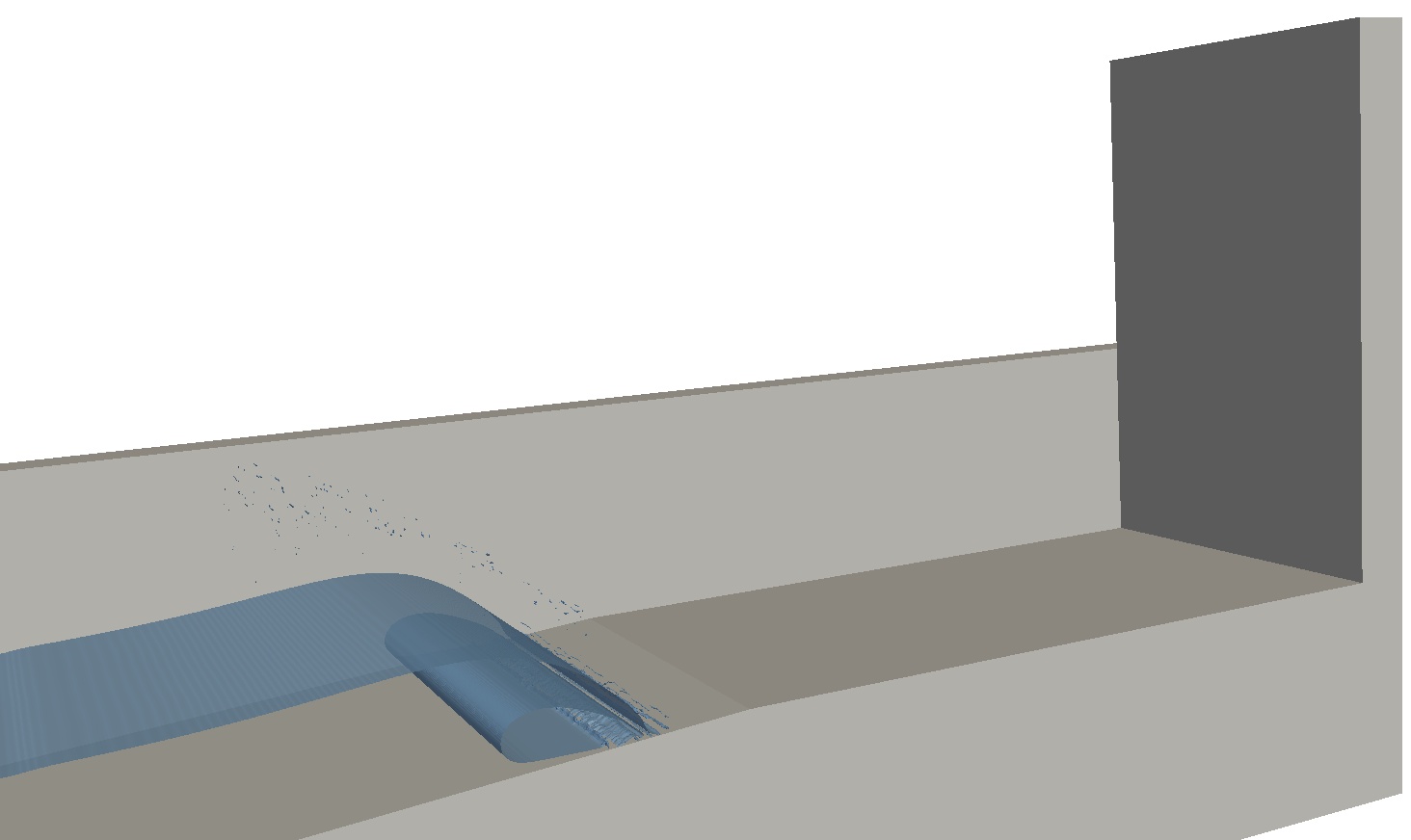}&\includegraphics[width=0.3\linewidth]{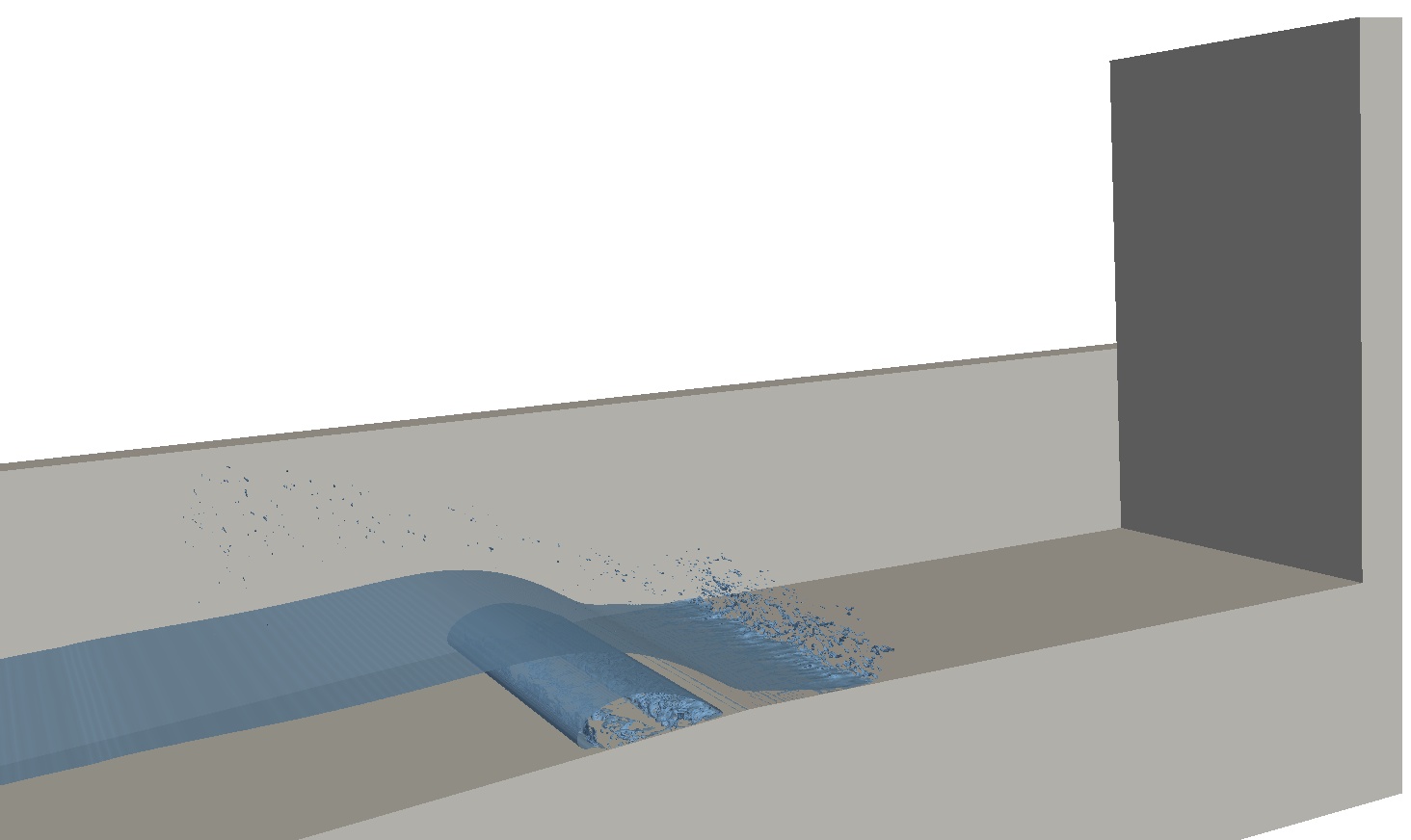}\\
(d) & (e) & (f)  \\
\includegraphics[width=0.3\linewidth]{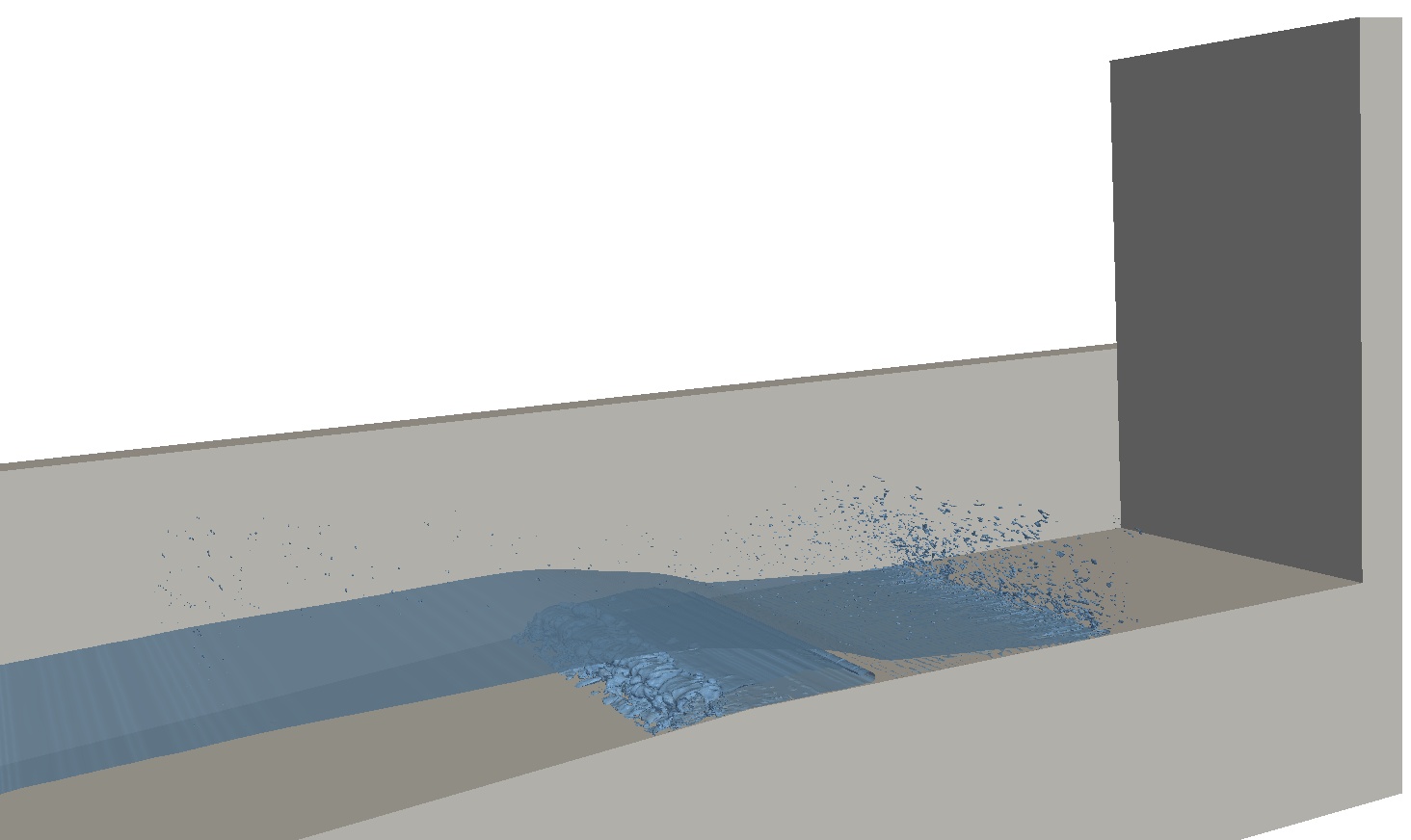}&\includegraphics[width=0.3\linewidth]{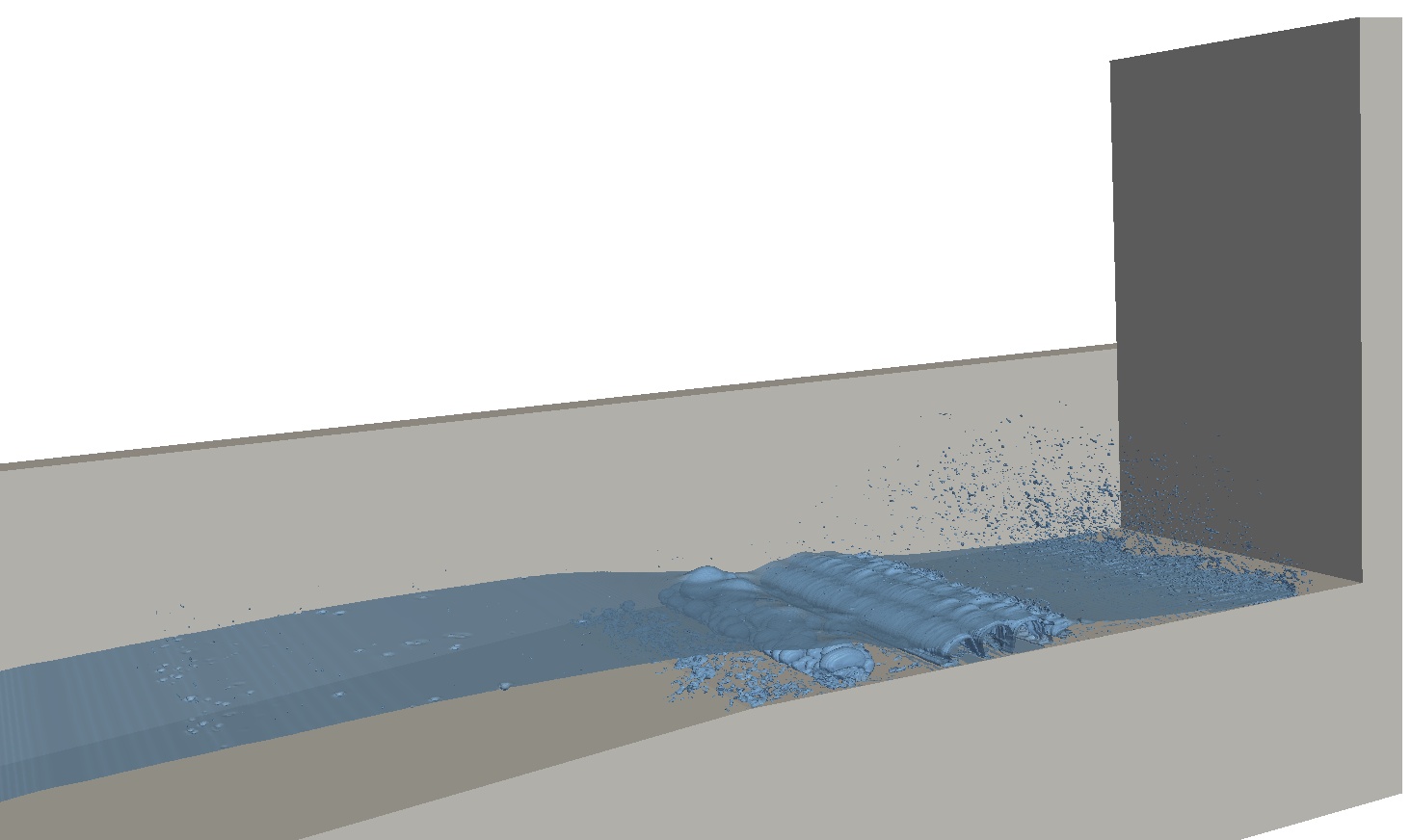}&\includegraphics[width=0.3\linewidth]{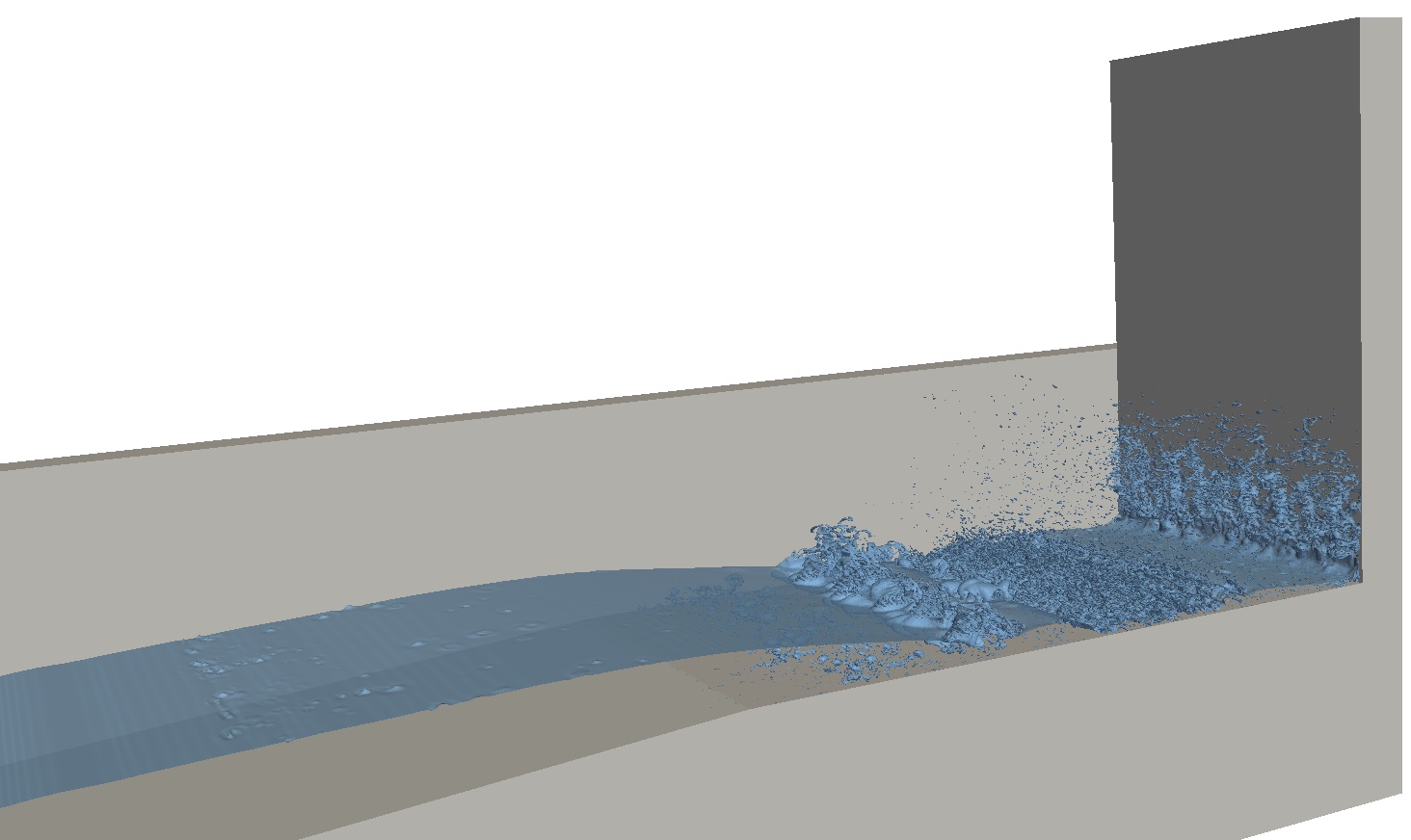}
\end{tabular}
\end{center}
\vspace*{-0.2in} \caption{NFA sequence showing a wave breaking on a beach and impacting breakwater.} \label{fig:nfa0}
\end{figure*} 

\begin{figure*}[h!t]
\begin{center}
\begin{tabular}{ccc}
(a) & (b) & (c) \\
\includegraphics[height=4.8cm]{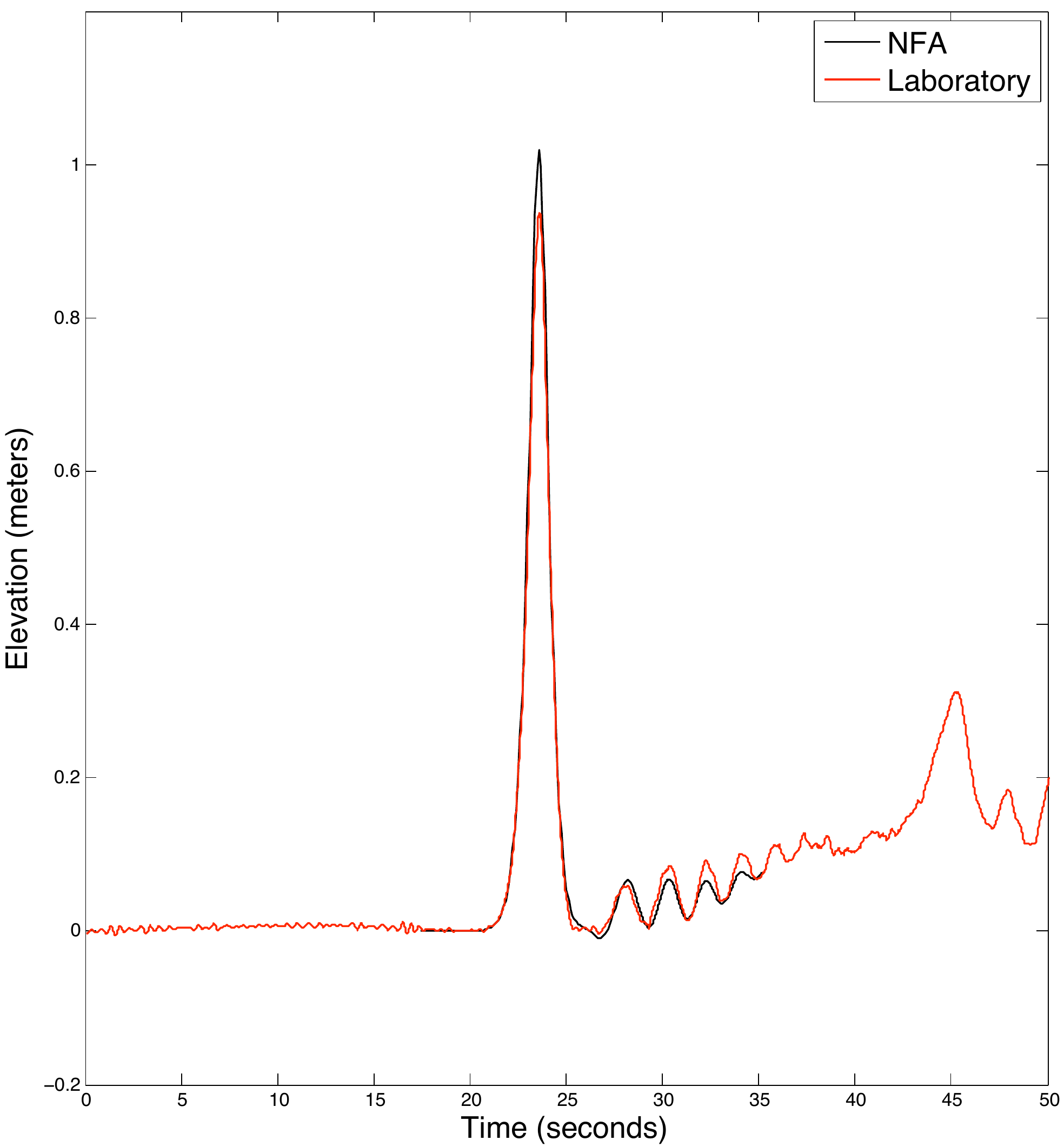}&\includegraphics[height=4.8cm]{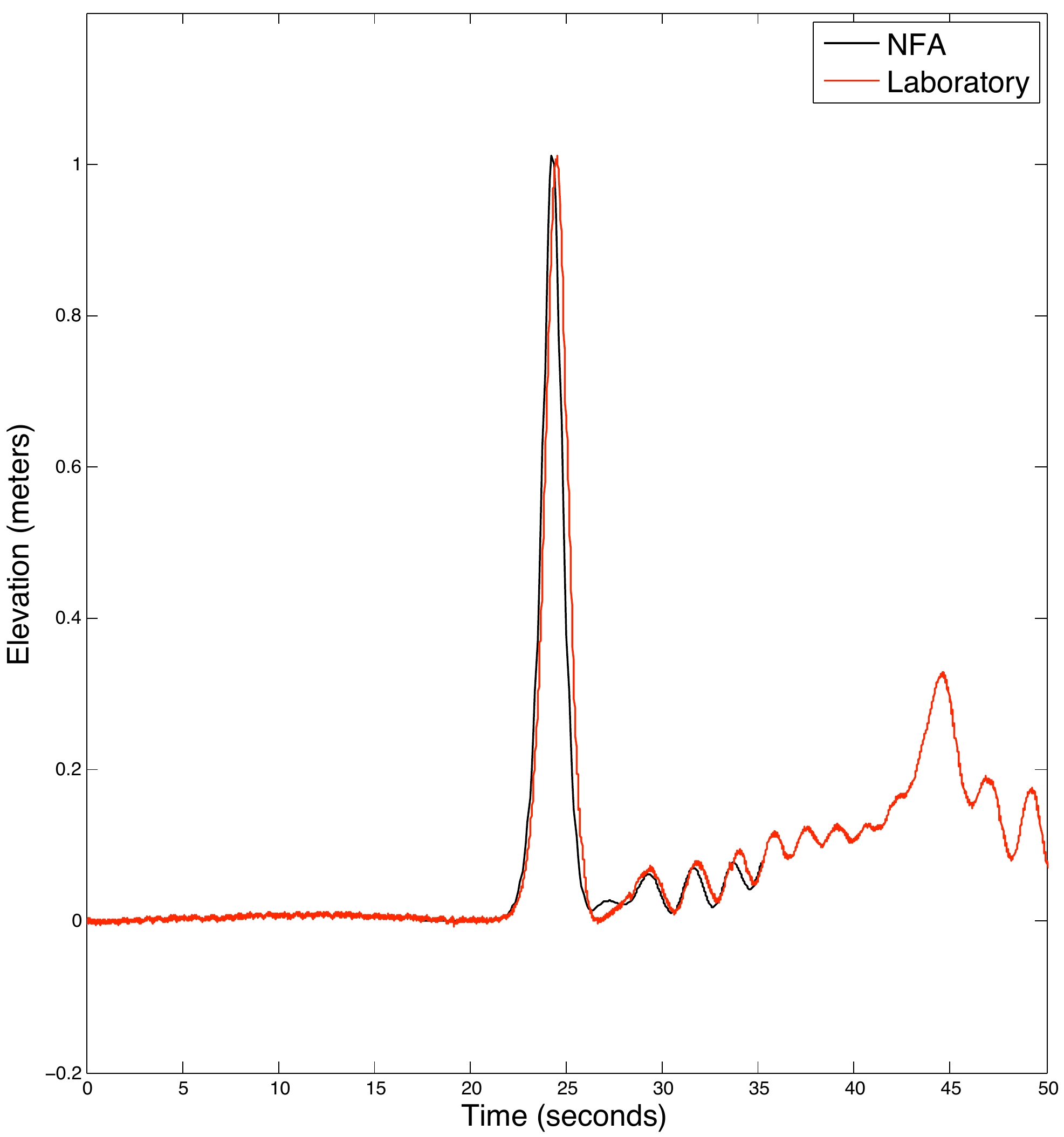}&\includegraphics[height=4.6cm]{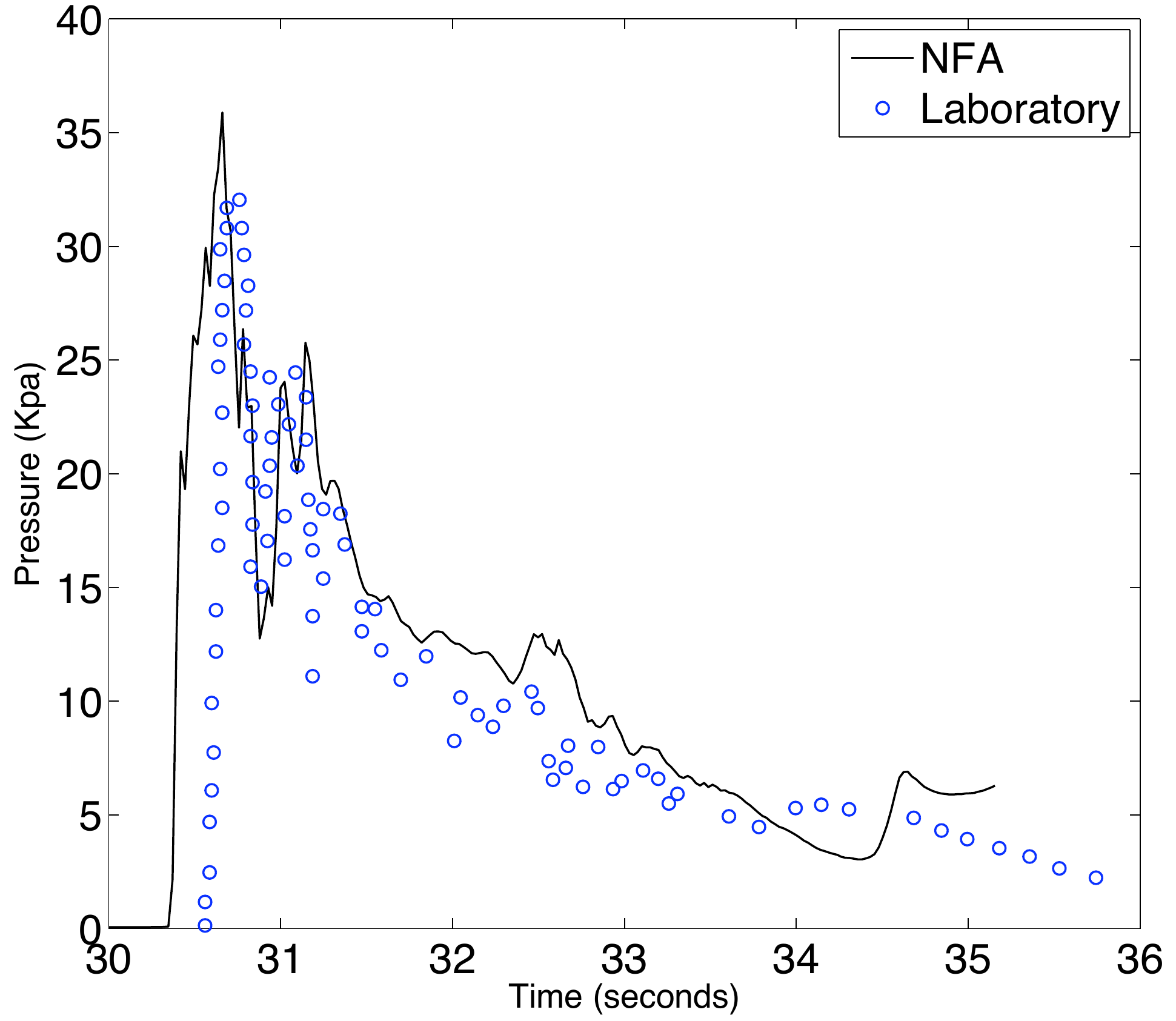} \\ \\
\end{tabular}
\end{center}
\vspace*{-0.5in} \caption{NFA predictions compared to wave tank measurements \citep{Oshnack09}. Wave probe data at (a) x=17.64m and (b) x=21.53m downstream of the wavemaker. (c) Pressure impact at base of breakwater at $(x,z)=(26.739,0.033)$. } 
\label{fig:nfa1}
\end{figure*} 

%\clearpage

In non-dimensional units, the extent of the computational domain is $-3 \leq x \leq 26.738$, $-0.80786\leq y \leq 0.80786$, and $-1.0 \leq z \leq 2.5$.   The grid spacing is stretched along the $x$ and $z-$axes.    The smallest grid spacing ($\Delta x = 0.00232$) along the $x-$axis is near where the soliton breaks ($20 \leq x \leq 23$) and near where the broken wave hits the breakwater ($25.5 \leq 26.73$).   The smallest grid spacing ($\Delta z = 0.00238$) along the $z-$axis is near the mean waterline ($z=0$).    The grid spacing is constant along the $y-$axis ($\Delta y = 0.00631$) . The number of subdomains along the $x$, $y$, and $z-$axes are respectively 36, 2, and 5.  Each subdomain has $128\times128\times128$ grid points.   The total number of grid points is 754,974,720.    The time step is variable.   In non-dimensional units relative to the time offset ($t=36$) that is used in the NFA simulation, the time step is $\Delta t=0.002$ for $0 \leq t \leq 23$; $\Delta t=0.0005$ for $23 < t \leq 24.75$; $\Delta t=0.00025$ for $24.75 < t \leq 25.5$; and $\Delta t=0.005$ for $25.5 < t \leq 36.75$.  The simulation is run for 40,500 time steps.   A 5-point density-weighted velocity smoothing scheme is applied every 20 time steps. The weights are $(1/8,1/4,1/4,1/4,1/8)$. The density ratio of air to water is $\lambda=0.001$.  The simulation takes 375 wall-clock hours using an older version of NFA with less optimization than the most recent version.  360 processors on the SGI Altix Ice at the U.S. Army Engineering Research and Development Center (ERDC) supercomputing center had been used to perform the simulation. 

Figure \ref{fig:nfa0} shows NFA predictions of the wave breaking on the beach, surging up the beach, and impacting the breakwater.   As shown in Figure  \ref{fig:nfa0}a-b, the soliton rises up the beach and overturns.   A pocket of air is trapped beneath the plunging breaker (Figure \ref{fig:nfa0}b-c). The air pocket breaks up due to a Kelvin-Helmholtz instability (Figure \ref{fig:nfa0}d-e). The formation of air pockets contributes to beach erosion due to scouring (Figure \ref{fig:nfa0}d-f).   A thin jet forms on the flat portion of the beach (Figure \ref{fig:nfa0}d-e).  The jet finally impacts the base of the breakwater (Figure \ref{fig:nfa0}f).

The NFA predictions are compared to laboratory measurements of the free-surface elevation at two locations down the flume (see Figure \ref{fig:nfa1}a\&b) and the impact pressure at the base of the breakwater (see Figure \ref{fig:nfa1}c). The free-surface elevations as predicted by NFA are in excellent agreement with the experimental measurements. This shows that NFA can simulate the propagation of waves over long distances with minimal amplitude and dispersion errors. The peak pressures predicted by NFA are also in good agreement with experiments, which is noteworthy because the impact loading occurred after the wave had broken and traveled up the beach after having traveled over 61m.  Pressure measurements show two distinct peaks (30.5s $\leq$ time $\leq$ 31.5s) corresponding to the cavity trapped by the plunging wave breaking up into two pieces (see
Figure \ref{fig:nfa0}e).  Pressures that are induced by the jet are important because in certain coastal areas buildings must be designed to sustain Tsunami loads.  

\section{Conclusions}

In this paper, two wave-impact scenarios are explored through experiments and numerical calculations: the impact of a deep-water plunging breaker on a partially submerged cube and the impact of run-up from a shoaling solitary wave on a vertical wall. The experimental measurements guide the development of numerical methods for predicting complex wave-impact problems.     Together, the experimental measurements and numerical predictions provide insight into the complex physics of wave-impact loading that would not otherwise be possible. In the case of the cube, a cube position that produces a strong vertical jet during a ``flip-through'' impact is examined with experiments and numerical calculations using the NFA code.  The experimentally measured and computed water surface profiles are found to be in close agreement.  Vertical accelerations of the contact point of the water surface on the front face of the cube are found to reach about 10$g$.    In the case of wave run-up on a vertical wall, NFA results are found to be in very good agreement with published experimental data, both in terms of water-surface height versus time and in terms of wall pressures versus time.

\subsection{Acknowledgements}

This work is supported in part by a grant of computer time from the DOD High Performance Computing Modernization Program \url{http://www.hpcmo.hpc.mil/}.  Animations
of NFA simulations are available at \url{http://www.youtube.com/waveanimations}.   We are grateful to the Office of Naval Research for funding this research. Dr.\ Ronald Joslin is the program manager.

\bibliography{29ONR}
\bibliographystyle{29ONR}

\end{document}